\documentclass[reprint,
superscriptaddress,
 amsmath,amssymb,
 aps,
 prl
]{revtex4-2}
\usepackage{graphicx}
\usepackage{xcolor}
\usepackage{dcolumn}
\usepackage{bm}
\usepackage{comment}
\usepackage{hyperref}
\usepackage[T1]{fontenc}
\usepackage{braket}
\usepackage{ulem}

\newcommand{\gtwo}[0]{g^{(2)}}
\newcommand{\ak}[1]{\hat{a}^{\phantom{\dagger}}_{#1^{\phantom{\prime}}}}
\newcommand{\akp}[1]{\hat{a}^{\phantom{\dagger}}_{#1^{\prime}} }
\newcommand{\adk}[1]{\hat{a}^{\dagger}_{#1^{\phantom{\prime}}} }
\newcommand{\adkp}[1]{\hat{a}^{\dagger}_{#1^{\prime}} }

\newcommand{\gdd}{g_{\mathrm{dd}}}
\newcommand{\fdd}{f_{\mathrm{dd}}}
\newcommand{\gamdd}{\gamma_{\mathrm{dd}}}

\begin{document}

\preprint{APS/123-QED}

\title{{Diminished quantum depletion and correlated droplets in one-dimensional dipolar Bose gas}}

\author{Bu\u{g}ra T\"uzemen}
\email{btuzemen@ifpan.edu.pl}
\affiliation{
Institute of Physics, Polish Academy of Sciences, Al. Lotnik\'{o}w 32/46, 02-668 Warsaw, Poland
}
\affiliation{
Center for Theoretical Physics, Polish Academy of Sciences, Al. Lotnik\'{o}w 32/46, 02-668 Warsaw, Poland
}

\author{Maciej Marciniak}
\thanks{Bu\u{g}ra T\"uzemen and Maciej Marciniak contributed equally to this work.}
\affiliation{
Center for Theoretical Physics, Polish Academy of Sciences, Al. Lotnik\'{o}w 32/46, 02-668 Warsaw, Poland
}

\author{Krzysztof Paw\l owski}
\affiliation{
Center for Theoretical Physics, Polish Academy of Sciences, Al. Lotnik\'{o}w 32/46, 02-668 Warsaw, Poland
}

\date{\today}

\begin{abstract}
We investigate the formation of self-bound states in a one-dimensional dipolar Bose gas under the influence of both strong short-range repulsive and strong nonlocal attractive interactions. While conventional methods such as the Bogoliubov–de Gennes (BdG) method typically fail in regimes with strong interactions due to significant quantum depletion, we reveal a particular scenario where the interplay of these strong interactions significantly mitigates quantum depletion, thus restoring the applicability of the BdG method. Remarkably, this restoration occurs even though the system exhibits pronounced antibunching, a feature usually linked with strongly correlated systems. By comparing our BdG results with exact ab initio calculations, we confirm the accuracy of the BdG approach in predicting the ground-state energy and correlation functions under these conditions. Furthermore, we demonstrate that adjusting the polarization direction of the dipoles allows for a tunable transition between a strongly interacting regime and the different balanced regime explored in this research.
\end{abstract}

\maketitle

\textit{Introduction.} Beyond-mean-field effects have received significant attention in the study of Bose-Einstein condensates due to their role in the formation of quantum droplets—a self-bound state stabilized by quantum fluctuations~\cite{Petrov_2015, Kadau_2016, Chomaz_2016, Ferrier_Barbut_2018, PhysRevLett.120.235301, PhysRevLett.120.135301, Bottcher_2019, PhysRevA.93.061603, PhysRevA.94.021602, PhysRevLett.119.255302}. These droplets emerge from a subtle balance between attractive and repulsive forces, where the mean-field interactions alone would typically predict either collapse or dispersion. In systems of atoms with magnetic moments, interactions include not only the short-range van der Waals potential, but also a long-range dipole-dipole potential. This interplay of interactions makes it essential to consider quantum fluctuations to accurately capture the system's physical behavior. Quantum fluctuations are incorporated into the Gross-Pitaevskii equation (GPE) through the Lee-Huang-Yang (LHY) correction, derived from Bogoliubov–de Gennes (BdG) analysis~\cite{lhy, SCH_TZHOLD_2006, Lima_2011, PhysRevA.86.063609}. This modification results in the extended Gross-Pitaevskii equation (eGPE), introducing a beyond-mean-field energy term which counteracts the possible collapse triggered by attractive interactions, thereby ensuring the droplet's stability.

\begin{figure}[t!]
\includegraphics[width=\linewidth]{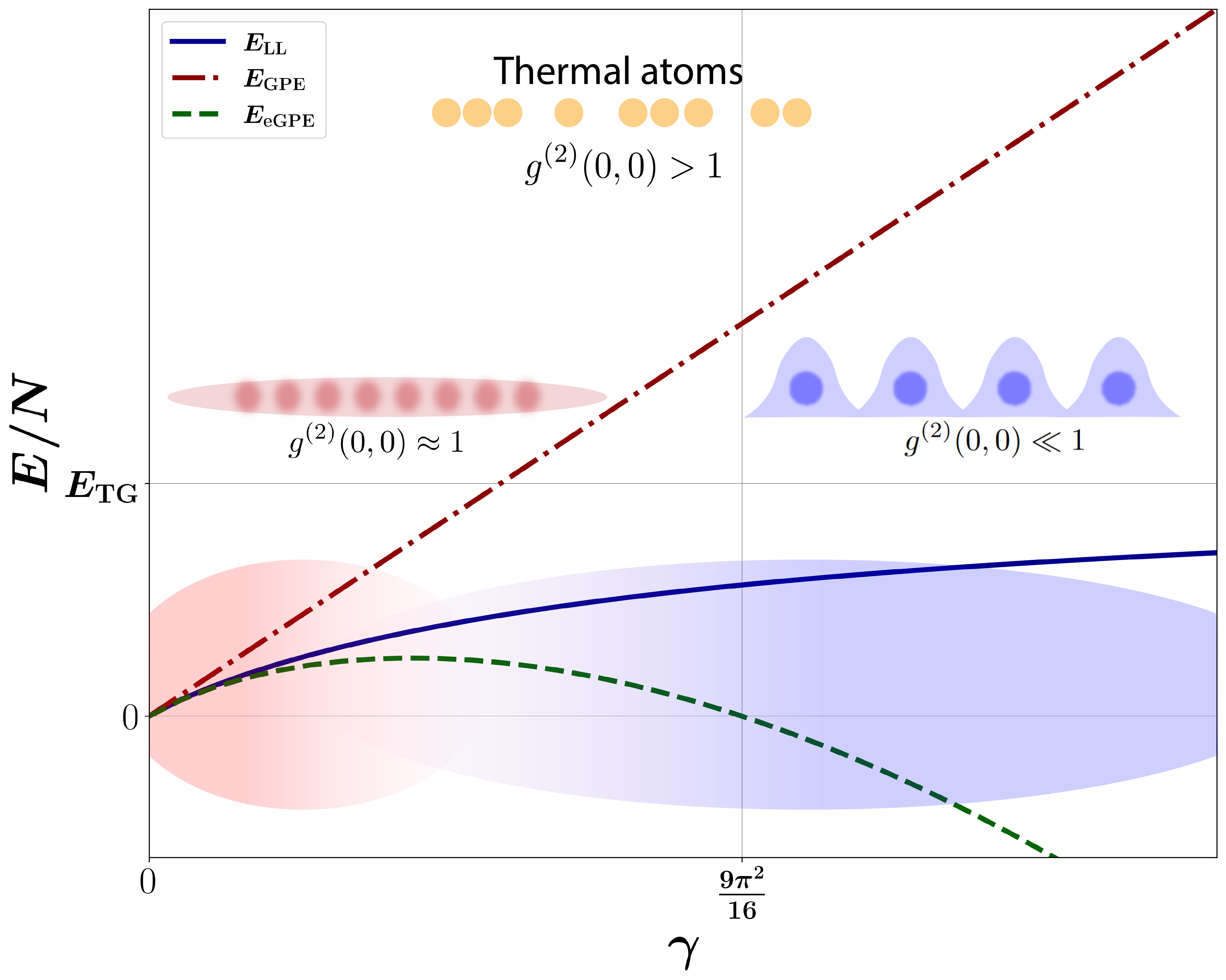}
\centering
\caption{ 
A comparison of the ground state energy in the Lieb-Liniger model per particle as a function of the Lieb parameter, $\gamma$ in the thermodynamic limit, obtained by different methods: exact Lieb-Liniger result, $E_{\mathrm{LL}}$  (solid line), the mean-field assuming a pure condensate, $E_{\mathrm{GPE}}$ (dot-dashed), and the energy corrected by the Lee-Huang-Yang term, $E_{\mathrm{eGPE}}$ (dashed). The tick labelled $E_{\mathrm{TG}}$ represents the energy in the Tonks-Girardeau limit. Additionally, the second-order correlation function, $\gtwo(x,x^\prime)$, is artistically depicted across different interaction regimes.}

\label{fig:graph_abs}
\end{figure}

The BdG analysis is crucial for evaluating quantum fluctuations. However, it relies on the assumption of low quantum depletion, making it mainly applicable in weakly interacting systems. As the interaction strength increases, the BdG approximation fails to reproduce the ground-state energy quantitatively. In one dimension (1D), it even predicts an unphysical negative-energy state for entirely repulsive systems (see dashed green line in Fig.~\ref{fig:graph_abs}).
This energy mismatch is directly related to the failure of BdG in predicting density-density fluctuations and increased quantum depletion for stronger interaction. In this study, we focus on these quantities and quantify the density fluctuations using the second-order correlation function $\gtwo$. It is usually used to study bunching properties, but also important in the context of the binary-interaction energy~\cite{Naraschewski_1999}, density fluctuations~\cite{Naraschewski_1999,PhysRevA.81.031610}, structure factor~\cite{PhysRevB.20.4912,Hung_2011} and even entanglement criteria~\cite{Dowling_2006,RevModPhys.80.517}.

Most of the studies on the systems with quantum fluctuations focus on the weakly interacting regimes in 2D~\cite{Norcia_2021, Bland_2022} and in 3D, where the droplet was demonstrated~\cite{Chomaz_2016, Schmitt_2016, Ferrier_Barbut_2018, Bottcher_2019, Modugno_2019}. The 1D droplets were discussed in the Bose-Bose mixture via Bogoliubov approach \cite{Petrov2016} or ab initio methods \cite{Parisi2019, Parisi2020, Kopycinski2023}. In 1D dipolar system, as shown in a series of publications, mostly based on phenomenological method~\cite{Oldziejewski_2020, baizakov_solitons_2009, dePalo2020, Edler_2017} or variational approaches \cite{dePalo2020}, a quantum droplet shall also form, even in the strongly interacting regime \cite{Oldziejewski_2020}. Although clear observations have yet to be made, current experiments on quasi-one-dimensional dysprosium atoms \cite{Kao2021Jan, kuanYu2023} and recent advancements with polar molecules in the ultracold regime \cite{Bigagli_2024} open new avenues for studying strongly correlated quantum liquids in the near future. This reduced geometry is particularly intriguing from both experimental and theoretical perspectives, as one can expect a significant reduction in particle losses (in 3D, which shortens the droplet's lifetime to a few milliseconds) and the emergence of strong correlations.

The 1D case has drawn considerable interest from both physicists and mathematicians, due to the Lieb-Liniger (LL) model~\cite{lieb, lieb2} for particles moving on a circle and interacting via a contact potential, $V(x) = g\delta(x)$. It is one of the few solvable but also useful models for interacting quantum systems. The key parameter in the LL model is Lieb's interaction parameter $\gamma = \frac{mg}{\hbar^2 n}$, where $n$ is the gas density. This parameter fully determines key quantities such as the ground-state energy per particle (see Fig.~\ref{fig:graph_abs}) and various correlation functions.
As $\gamma$ increases, the $\gtwo$ function evolves from a position-independent profile for a non-correlated system at $\gamma \sim 0$ through a regime of weak anti-bunching to complete anti-bunching in the Tonks-Girardeau limit $\gamma \to \infty$~\cite{Girardeau1960}. In fact, in a system with contact interactions only, the anti-bunching remains small only in a weakly interacting limit if  $\gamma \ll 1$, within the validity range of the BdG approximation.

In this Letter, we observe that due to the interplay of strong interactions, the coherence may be restored. Although the nonlocal and contact interactions act on different length scales, they still partially compensate for each other. The resulting state of the matter can have small quantum depletion, as in the weakly interacting regime, but exhibits substantial correlation (antibunching) due to the presence of high-momenta states. The strongly interacting system dominated by short-range repulsion was already studied in Ref.~\cite{Oldziejewski_2020} where the dipolar interaction was treated only as a perturbation. This work explores a different regime, where short-range repulsion and dipolar attraction are comparable in magnitude. Consequently, the Lieb-Liniger model is no longer an adequate starting point, requiring the use of the BdG approach, as demonstrated in this study. While strong antibunching was reported in Ref.~\cite{Oldziejewski_2020}, this work finds weaker antibunching in the considered regime, albeit with notable agreement between BdG predictions and the \textit{ab initio} results.

We demonstrate that the BdG approach provides a qualitatively reliable framework for predicting self-bound states, such as quantum droplets, even across a broad range of interaction strengths. For 1D gases and self-bound states, our findings show remarkable agreement between number-conserving Bogoliubov analysis and exact ab initio calculations, particularly in regimes where strong interactions are nearly balanced by attractive interactions.

\textit{The model.} We analyze a system comprising $N$ dipolar bosons polarized along $x$ direction in a quasi-1D geometry with periodic boundary conditions. The quasi-1D geometry may be achieved by imposing a tight harmonic trap with high frequency $\omega_{\perp}$ in the transverse directions, $y$ and $z$. We assume that in these directions, the bosons' wavefunction is well-approximated by a Gaussian profile. In the longitudinal direction $x$, the system has a finite length $L$. To get an effective dipolar interaction in quasi-1D, we integrate the full 3D interaction potential over the transverse directions~\cite{PhysRevA.81.063616, PhysRevA.87.039903}, assuming that every particle is in the single-particle ground Gaussian state in the $y-z$ plane. Finally, the effective interaction potential $V_{k}$ in momentum space,  including both short-range and long-range components, takes the form: 
$V_{k} = g - \gdd\left(1 - f\left(\frac{k^2 \sigma^2}{2}\right)\right)$,
where $\gdd = \frac{\mu_{0}\bm{d}^2}{2\pi\sigma^2}$ is the dipolar coupling constant, $\sigma = \sqrt{\frac{\hbar}{m\omega_{\perp}}}$ represents the oscillator length in the transverse directions, $\bm{d}$ is the atomic dipole moment, $\mu_{0}$ is the vacuum permeability, and $f(u) = ue^{u}\mathrm{E}_1(u)$ with $\mathrm{E}_1(u)$ being the exponential integral function.  

The many-body Hamiltonian of our quasi-1D system, in the momentum representation, is expressed as:
\begin{equation}
\mathcal{H} = \sum_{k} h_{k} \adk{k} \ak{k} + \frac{1}{2L}\sum_{k,k^\prime,q} V_{q} \adk{k+q}\adkp{k-q}\akp{k}\ak{k}, 
\label{eq:ham1}
\end{equation}
where $h_{k} = \frac{\hbar^2 k^2}{2m}$ is the single-particle kinetic energy, and the operators $\ak{k}$ and $\adk{k}$ obey the bosonic commutation relations $[\ak{k},\adkp{k}]=\delta_{k,k^\prime}$. The summation goes over discrete momenta $k=2\pi j/ L$, for an integer $j$.

The simplest approximation to the Hamiltonian given in Eq. \eqref{eq:ham1} is the Gross-Pitaevskii (GP) approach, which assumes that all atoms occupy the same orbital. We focus on the case of the short-range repulsion exceeding the attractive interaction, so the GP approach gives a constant wave function as the condensate~\footnote{We define the term condensate as referring to the $k=0$ mode.} orbital with energy per particle equals to $E_{\mathrm{GPE}}/N =  \frac{1}{2} n V_0 $, where $n$ is the gas density. This GP energy fails shortly as the interactions get stronger (see Fig.~\ref{fig:graph_abs}). Although the single-particle density remains uniform even for interacting systems, the interactions imply the correlation between atoms, leading to the depletion of the condensate orbital.
As long as this quantum depletion is small, one can employ the Bogoliubov method, where the interaction between the depleted atoms and the condensate is taken into account perturbatively. The Bogoliubov method can be used to study not only the corrections to the GP ground state energy but also to obtain a better approximation to a ground state wavefunction, the correlation functions, the quantum depletion, and the elementary excitations of the ultracold Bose gas. 

Here, to compare with exact few-body results, we use the number-conserving BdG version ~\cite{Leggett_2001}, where the ground state reads:
\begin{equation}
\ket{\Psi_{\mathrm{BdG}}} = C \left( \hat{a}_0^{\dagger}\hat{a}_0^{\dagger} - \sum_{k \neq 0}^{\infty} \frac{v_k}{u_k} \hat{a}_k^\dagger \hat{a}_{-k}^\dagger \right) ^{N/2} \ket{{\rm vacuum}}.  
\label{eq:bdg_ansatz}
\end{equation}
Here, the Bogoliubov amplitudes, assuming  uniform single-particle density of a condensate, are $|u_{k}|^2 = 1 + |v_{k}|^2 = (1/2)\left[ \left(h_{k} + n V_{k}\right)/\varepsilon_{k} + 1 \right]$, $\varepsilon_{k} = \sqrt{ h_{k}\left(h_{k}+2nV_{k} \right)}$,
and $C$ is a normalization constant.

For stronger interactions, the Bogoliubov approach usually leads to unphysical results. For large $\gamma$, one must use the {\it ab initio}~\footnote{ By {\it ab initio} calculation, we mean to find a direct solution of the model, which is many-body but already one dimensional and with effective potentials for dipolar and short-range interactions.} model~eqref{eq:ham1}, which is analytically solvable for nondipolar cases, but must be handled numerically otherwise. In this Letter, depending on the system size we use exact diagonalization (number of particles $N < 6$) or the Density Matrix Renormalization Group for Continuous Quantum Systems (cDMRG) \cite{Dutta2022Jun} (for systems with $N = 6$). More details about cDMRG are given in Supplemental Material.

Below, we show our results for the ground state, using all the approaches mentioned above. In what follows, we use the box units, with $L$, $mL/\hbar$, and $\hbar^2/mL^2$ as the units of length, time, and energy, respectively. To characterize the interaction regimes, we use Lieb's parameter and its counterpart $\gamdd=\gdd/n$ for dipolar interaction.

\textit{Energy profiles and the self-bound states} 
Our study centres on the analysis of the ground state energy. For a non-dipolar case, $\gamma_{\rm dd}=0$, the model is just the Lieb-Liniger model. 
For large $\gamma$, the energy from the exact LL model increasingly deviates from the mean-field and BdG predictions, as illustrated in Fig.~\ref{fig:graph_abs}. The ground state energy obtained via BdG approximation in the thermodynamic limit equal to $E_{\mathrm{eGPE}}/N=\frac{1}{2}gn - \frac{2\sqrt{m}}{3\pi\hbar}g\sqrt{gn}$, leads to unphysical result - the negative energy for an entırely repulsive system for $\gamma > \frac{9\pi^2}{16}$. The reason behind the mismatch is the assumption underlying the BdG method -- the number of particles occupying the $k\neq 0$ states must be negligible compared to those occupying the $k=0$ state. In reality, for larger parameter$\gamma$, the ground state spans over many momentum modes. Eventually, for $\gamma\to\infty$, the ground state energy converges to a constant, $E_{\rm TG}/N = \frac{\pi^2 \hbar^2 n^2}{6m}$ in the Tonks-Girardeau limit~\cite{Girardeau1960}. 

\begin{figure}[t!]
\includegraphics[width=\linewidth, keepaspectratio, trim=0cm 0cm 0cm 0cm, clip]{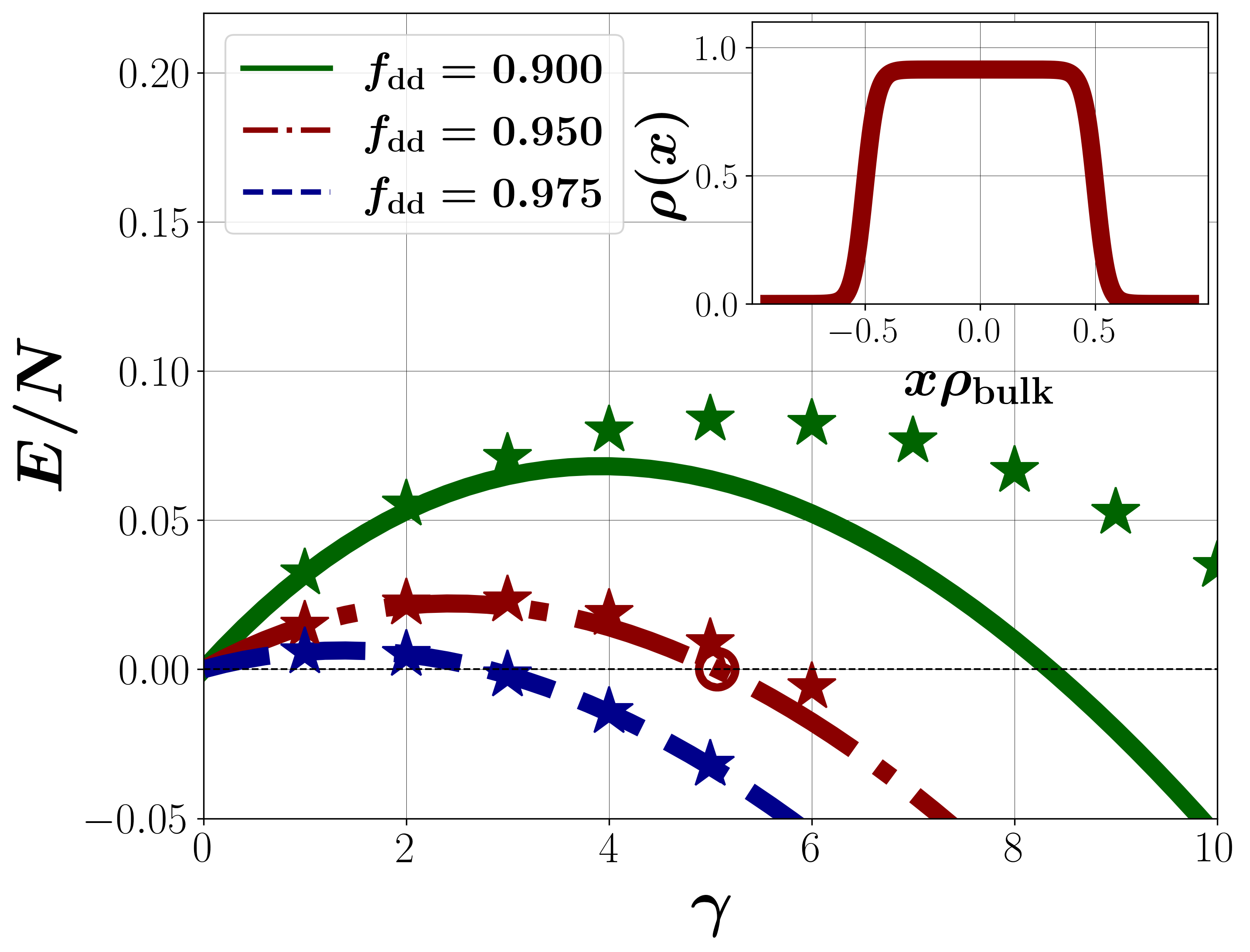}
\centering
\caption{The energy per particle as a function of $\gamma$ calculated using the BdG method (lines), and exact diagonalization energies (stars) for various strengths of dipolar interaction. Here, the number of particles $N=4$, and $n\sigma=0.05$, where $n$ is the particle density. The inset depicts the density profile for $\gamma = 5$ and $f_{\mathrm{dd}} = 0.95$ (marked by the circle), obtained from the BdG analysis in the main figure, based on Eq. \eqref{eq:eGPE} but for $N=34$ atoms.}
\label{fig:energies}
\end{figure}

Once the dipolar interaction is included, the ground state energy may change qualitatively. For fixed ratio $\fdd=|\gdd/g|$, when $\gamma\to\infty$ the ground state energy converges to  $-\infty$, indicating the emergence of self-bound states. The origin of this phenomenon is simple~\cite{Oldziejewski_2020}: for a large $\gamma$ the system becomes anti-bunched, resulting in a diminished contact interaction energy. Simultaneously, the non-local attraction persists, keeping all atoms together. As a result, the atomic cloud may be broad and diluted. When the system is converging towards a fermionized state in the Tonk-Girardeau regime \cite{Girardeau1960}, then its kinetic energy scales as $n^2$ with the density, it remains lower than the non-local interaction energy,  scaling like $n$. Eventually, at a critical value $\gamma_c$, the total energy becomes negative, leading to the formation of a self-bound state.
We illustrate this in Fig.~\ref{fig:energies} showing the exact values of the ground state energies (solid lines) for various fixed ratios $\gamma/\gamma_{\rm dd}$. As one may expect, the stronger non-local attraction, the quicker the system becomes self-bound.

In the strong interaction regime, the exact ground-state energy of the dipolar system approaches ${E_{\rm GS}\to-\infty}$, similar to what is observed in the BdG approximation for non-dipolar systems when applied beyond its validity range. This raises the question of whether the negative energy predicted by the BdG approximation for dipolar systems is physically meaningful or remains unphysical. To address this, we investigate the issue in detail, starting with the BdG energy~\cite{lhy}. In few-body systems, to obtain the ground-state energy, it is necessary to use the ground-state ansatz at Eq.~\ref{eq:bdg_ansatz} on Eq.~\ref{eq:ham1}. As a result, one gets \footnote{See Supplemental Material for details.}:

\begin{eqnarray}
E_{\mathrm{GS}} &=& \frac{1}{2}V_0 \left[ N(N-1) |C_0|^2 + (N-2)(N-3)(1 - |C_0|^2) \right] \nonumber \\
&+& 2 \sum_{k>0} \left( \left(h_k + (N-2)(V_{k} + V_0) \right) |C_k|^2 \right) \nonumber \\
&+& 2 \sum_{k>0} \sqrt{N(N-1)} C_k^* C_0 V_k,
\end{eqnarray}
where $C_{0} = C\sqrt{N!}$, and $C_{k} = -\frac{v_{k}}{2u_{k}}CN\sqrt{\left(N-2\right)!}$. By numerically evaluating the above equation, we obtain the quantum corrections to the mean-field energy. Fig.~\ref{fig:energies} illustrates the comparison between energies calculated using the BdG method and exact diagonalization. Notably, the agreement between the Bogoliubov approximation and the exact results improves as the ratio of dipolar to contact interaction strengths, $\fdd$, approaches $1$, even in the case of negative energies.

Due to this agreement, the BdG method becomes suitable for determining a critical interaction parameter, $\gamma_{c}$, at which the system's energy reaches zero. For a given $\fdd$, the value of the $\gamma_{c}$ mainly depends on the range of the dipolar interaction, $\sigma$. In Fig.~\ref{fig:energies}, we present results for a dilute system with $n \sigma=0.05$. For any finite $\sigma$, however small, the self-bound state will emerge, but they do not exist in the limit of vanishing non-locality  $\sigma\to0$. There, the Hamiltonian transitions the LL model with the interaction strength $(g-\gdd)$. It should be noted that the BdG analysis incorporates beyond-mean-field corrections but assumes a very low level of quantum depletion (in other words, a very large condensate fraction) by definition. As a result, the increased energy contribution due to higher depletion is not captured in the BdG approximation. Although the overall depletion may still be relatively small, it is enough to cause a discrepancy between the BdG approximation and the exact results, leading to an underestimation of the total energy.

Apparently, for a dipolar system, the negative BdG energy is meaningful and indicates the emergence of a self-bound state. To show that, we use the numerical results for the energy of a small and homogenous system as a function of the density $E[\rho]$ written in the form 
\begin{equation}
  E[\rho]=  \int dx \psi^*(x)\mathcal{H}_{\rm eGPE}[\psi^*(x), \psi(x)] \psi(x)
  \label{eq:eGPE}
\end{equation} 
for the fields $\psi(x)$ and an extended GPE Hamiltonian $\mathcal{H}_{\rm eGPE}$. Technically, in the limit of a large number of atoms, this shall coincide with a typical procedure with the LHY corrections added to GPE. In our case, we use the total energy function (inferred from a small system), with appropriate Bogoliubov modes and in the regime, where we benchmarked it with exact diagonalization. We then use this function for slightly larger systems. 
As an example of a bound state, in the inset of Fig.~\ref{fig:energies}, we provide the density profile of a droplet obtained by solving the GPE extended by the numerically obtained quantum correction. 

\begin{figure}[t!]
\includegraphics[width=\linewidth, keepaspectratio, trim=0cm 0cm 0cm 0cm, clip]{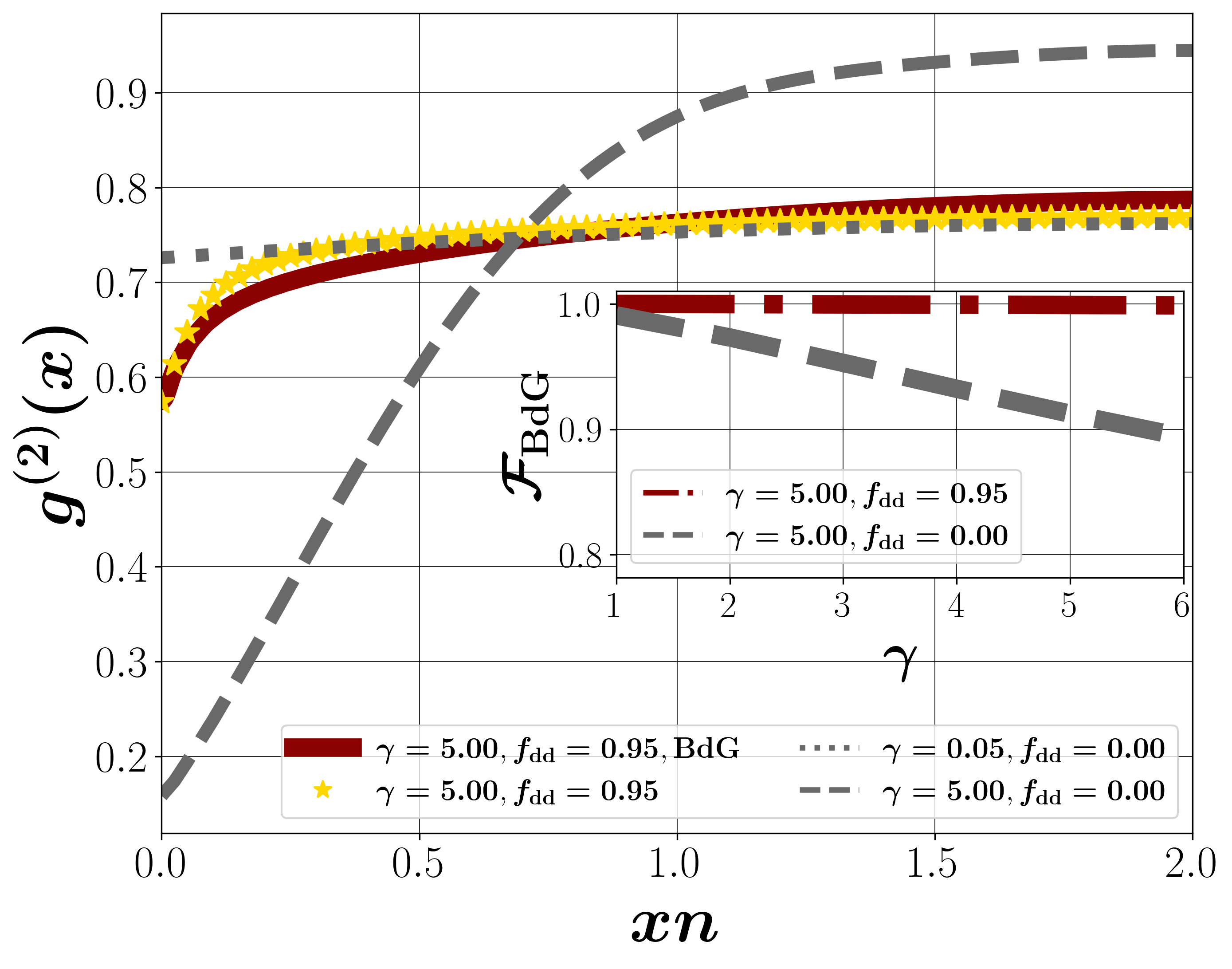}
\centering
\caption{The second-order correlation function, $\gtwo(x)$. We present the two limiting cases, $\gamma = 0.05$ (dotted line) and $\gamma=5$ (dashed line), for the non-dipolar case. The remaining two lines—starred and solid—correspond to the dipolar case with $\gamma = 5$ and $f_{\rm dd} = 0.95$, obtained via exact diagonalization and the BdG approximation, respectively. The solid line corresponds to the circled point in Fig.~\ref{fig:energies} and represents the expected correlation function for a droplet presented in the inset of Fig.~\ref{fig:energies}. Inset shows the fidelity between the wave functions coming from the ab initio methods and the BdG for two characteristic ratios $f_{\rm dd}$. Here, the number of particles $N=4$, and for the dipolar case, $n\sigma=0.05$.}
\label{fig:g2}
\end{figure}

\textit{Second-order correlations and the quantum depletion}
The observation that the BdG method performs surprisingly well in predicting the ground state energy raises the question of whether this agreement is a coincidence or if there are deeper reasons behind it. To answer this, we study the many-body properties of the ground state, using {\it ab initio} exact diagonalization treatment and the number-conserving Bogoliubov equations. We begin with the examination of the second-order correlation function, defined as
$g^{(2)}({x}) = \frac{\braket{\hat{\psi}^\dagger({x}) \hat{\psi}^\dagger(0) \hat{\psi}(0) \hat{\psi}({x}})}{\braket{\hat{\psi}^\dagger({x}) \hat{\psi}({x})} \braket{\hat{\psi}^\dagger(0) \hat{\psi}(0)}}$, where $ \hat{\psi}({x})$ is a bosonic field operator. As illustrated in Fig.~\ref{fig:g2}, in the non-dipolar case, but with relatively large contact interaction, $\gamma=5$ 
the second-order correlation function is diminished
$\gtwo(0)\ll 1$ highlighting the system's strongly interacting nature. When strong dipolar interactions are introduced, one may expect that the strong but competing contact and dipolar interactions cancel each other out, bringing the system back to a weakly interacting regime. Contrary to this expectation, our correlation analysis reveals enhanced anti-bunching (compare with a dotted line in Fig.~\ref{fig:g2}), which arises from the localized nature of the strong contact interaction. We have checked the validity of the BdG method by computing fidelity between the numerically obtained ground state of the Hamiltonian \eqref{eq:ham1} and the BdG Ansatz \eqref{eq:bdg_ansatz}, $\mathcal{F}_{\mathrm{BdG}} = |\langle \Psi_{\mathrm{exact}} | \Psi_{\mathrm{BdG}} \rangle |^2$ (see the inset of Fig.~\ref{fig:g2}). We report that, in cases with strong dipolar interactions, this fidelity is notably high, affirming the accuracy of the BdG ground state ansatz. Leveraging this accuracy, we also computed $\gtwo(x)$ using the BdG ansatz and obtained a profile that closely matches the one derived from {\it ab initio} methods. This explains the agreement between energies shown in Fig.~\ref{fig:energies}. The $\gtwo(x)$ profile in Fig.~\ref{fig:g2}, for the dipolar case, corresponds to the droplet shown in Fig.~\ref{fig:energies}, which means that the droplet formed in the dilute limit exhibits a significant second-order correlation, albeit with a less pronounced, yet evident, degree of anti-bunching.

\begin{figure}[h!]
\includegraphics[width=\linewidth, trim=0 0 0cm 0cm, clip]{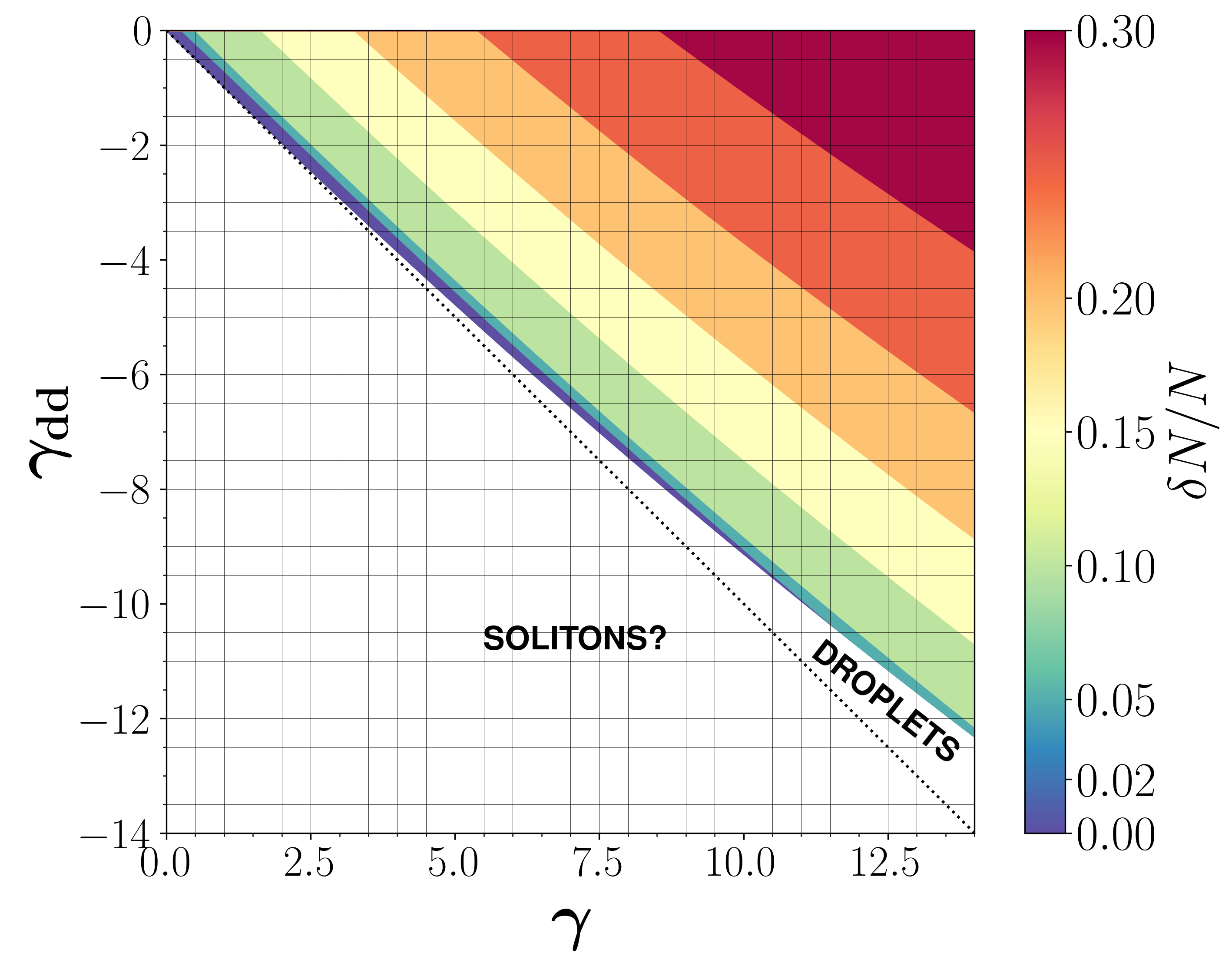}
\centering
\caption{ The quantum depletion, $\delta N/N$ as a function of contact and dipolar interaction strengths. Here, the number of particles $N=4$, and $n\sigma=0.05$. The data was obtained by exact diagonalization.
}
\label{fig:depl}
\end{figure}

The above discussion indicates that there exists a regime of nearly balanced interactions where the assumption of low depletion, as used in the BdG model, is sufficiently accurate. This is further supported by examining quantum depletion

\begin{equation}
    \delta N = \sum_{k\neq 0}  \langle{\hat{a}_k^\dagger\hat{a}_k}\rangle
\end{equation}

across the parameter space as shown in Fig. \ref{fig:depl}, which illustrates how depletion varies with different strengths of contact and dipolar interactions. Notably, along the crossover line between gaseous and liquid states, there is a narrow region where quantum depletion remains low. For $n\sigma=0.05$, it is observed that along the transition line up to $\gamma\approx 12$, depletion is low, suggesting that the BdG approach is reliable in this region. The low-depletion region is finite and disappears at large $\gamma$.
However, as $\sigma n$ decreases, the BdG model remains applicable even at higher interaction strengths. Additionally, in the case where $|\gdd|>|g|$, one expects a crossover from droplet to bright soliton -- this region is marked with a question mark, as a topic for further study ~\cite{Oldziejewski_2020}.

\textit{Possible experimental demonstration}
The experimental realization of this study presents two potential challenges. First, the dipolar strength of atoms is typically fixed, meaning that adjusting $\gamma_{dd}$ can only be achieved by varying the density. However, recent advancements in ultracold dipolar molecules~\cite{Bigagli_2024} suggest that the dipolar strength may soon be adjustable. Second, creating a system that exhibits true 1D characteristics is challenging. In related experiments, dysprosium atoms with magnetic moments of $|\bm{d}|= 10\mu_{\rm B}$, where $\mu_{\rm B}$ is the Bohr magneton, were confined in quasi-1D harmonic traps, with a longitudinal-to-transverse oscillator length ratio of approximately $l_{\parallel}/l_{\perp}\approx25$, and an average atom count per trap of $N\approx20$ yielding to an effective dipolar interaction range ${\color{red}n} \sigma \approx 1$~\cite{Kao2021Jan}. To observe the predicted low depletion and anti-bunching effects, shorter effective ranges of dipolar interactions are necessary. These conditions can be met within current experimental setups by increasing the transverse trap frequencies or decreasing gas density. For example, raising the transverse trap frequencies to $2\pi \times 100$ kHz and reducing the average number of atoms per trap~\cite{Li2023Jun} to $N\approx 6$, could bring the effective interaction range down to $n \sigma \approx 0.1$, making the observation of the anticipated effects feasible.

\begin{figure}[h!]
\includegraphics[width=\linewidth]{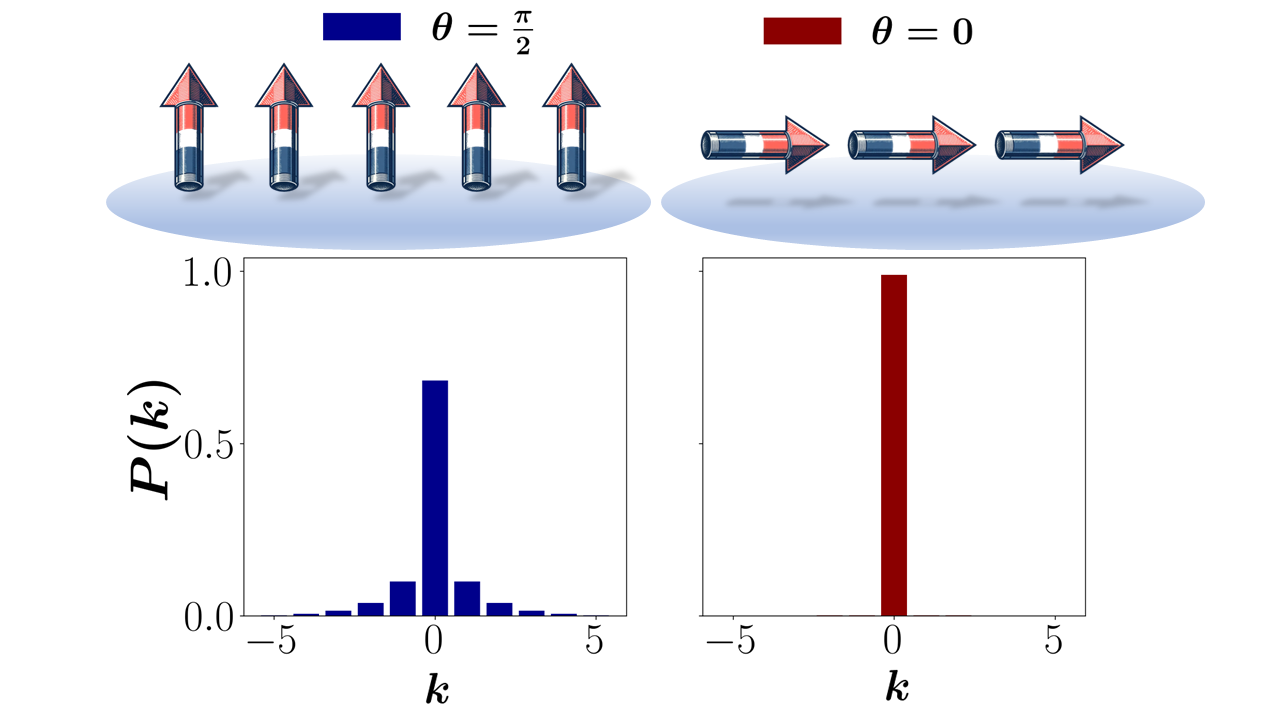}
\centering
\caption{ The momentum distribution for two particular dipolar orientations obtained by using cDMRG. In the calculations, a model comprising $N_S=60$ segments was used, with a cutoff range for non-local interactions set to $r=10 n\sigma$. For more details, see supplemental materials. Here we consider the system with $N=6$ atoms and effective interaction range $n \sigma = 0.1$, $\gamma = 6.7$, $\fdd = 0.88$. The effective range of dipolar interactions and the number of atoms correspond to the estimated limitations of current experimental setups~\cite{Kao2021Jan,Li2023Jun,Yang2023Aug}. The interaction strengths, $\gamma$ and $\gamma_{dd}$ were chosen such that for $\theta = \frac{\pi}{2}$ (left panel), the system behaves as a strongly repulsive one, while for $\theta = 0$ (right panel), it behaves as a nearly self-bound state, close to the transition to the liquid phase, with low depletion.
}
\label{fig:exp}
\end{figure}

In Fig.~\ref{fig:exp}, we present a simple experimental concept to demonstrate the results of this Letter. Achieving a stable system and accessing the narrow, low-depletion region may be challenging. However, it would be easier to enter the strongly interacting regime by arranging the dipoles in a side-by-side configuration, where all interactions are repulsive, as shown in the left panels of Fig.~\ref{fig:exp}. By tuning the interaction, one could transition between the strongly interacting, TG-like regime,  and the Bose-Einstein condensate regime with almost all atoms occupying the $k=0$ mode as depicted in the right panel of Fig.~\ref{fig:exp}. 

Our cDMRG results with $N=6$ particles indicate a significant difference in momentum distributions, particularly the high occupation of the $k=0$ mode in the latter case. The occupation of the $k=0$ mode, associated with the condensate, serves as a direct indicator of reduced depletion and validates the BdG approximation discussed earlier. Moreover, from an experimental standpoint, this difference in momentum distributions should be even more pronounced in a real setup with a larger number of particles.

In conclusion, we have demonstrated that the dipolar atoms in 1D can form a self-bound state, a quantum droplet, that locally exhibits substantial anti-bunching while remaining a coherent matter wave. The anti-bunching observed at short distances originates simply from strong short-range repulsion, whereas the non-local attraction is sufficient to restore coherence.  
Although in the non-dipolar case, for $\gamma \geq 1$, the Bogoliubov de Gennes approximation fails, giving even the negative ground state energy in the entirely repulsive system, in the dipolar case the BdG is accurate even for large values of $\gamma$. The negative energy, as predicted via BdG, agrees with the exact {\it ab initio} even in the regimes with droplets. We support the statement investigating energy, quantum depletion, momentum distribution, and the second-order correlation function using {\it ab initio} methods and the number-conserving Bogol{\color{red}i}ubov approximation.

Having this observation in hand, we formulated an idea for the experiment -- in principle,  the ground state coherence may be steered in the dipolar system, from the Tonks-Gireardeau regime to a fully coherent matter wave, by simply tuning the dipole polarization direction.
Noteworthy, the strongly interacting 1D regime was achieved experimentally \cite{Haller2009Sep, Kao2021Jan}, although it is still debated which of the 1D interaction regimes can be achieved for dipoles \cite{Zin2021Sep, Pylak2024}  -- this will be the subject of our future research. Furthermore, a natural extension of this study involves exploring the excitation spectra of these bound states and investigating the transition from droplets to bright solitons, which we plan to address in subsequent work.

\acknowledgments
We thank G. Astrakharchik for his interesting comments. B.T., M. M., and K.P. acknowledge support from the (Polish) National Science Center Grant No.
2019/34/E/ST2/00289. Center for Theoretical Physics of the Polish Academy of Sciences is a member of the National Laboratory of Atomic, Molecular, and Optical Physics (KL FAMO).

\bibliography{biblio}

\begin{thebibliography}{50}%
\makeatletter
\providecommand \@ifxundefined [1]{%
 \@ifx{#1\undefined}
}%
\providecommand \@ifnum [1]{%
 \ifnum #1\expandafter \@firstoftwo
 \else \expandafter \@secondoftwo
 \fi
}%
\providecommand \@ifx [1]{%
 \ifx #1\expandafter \@firstoftwo
 \else \expandafter \@secondoftwo
 \fi
}%
\providecommand \natexlab [1]{#1}%
\providecommand \enquote  [1]{``#1''}%
\providecommand \bibnamefont  [1]{#1}%
\providecommand \bibfnamefont [1]{#1}%
\providecommand \citenamefont [1]{#1}%
\providecommand \href@noop [0]{\@secondoftwo}%
\providecommand \href [0]{\begingroup \@sanitize@url \@href}%
\providecommand \@href[1]{\@@startlink{#1}\@@href}%
\providecommand \@@href[1]{\endgroup#1\@@endlink}%
\providecommand \@sanitize@url [0]{\catcode `\\12\catcode `\$12\catcode `\&12\catcode `\#12\catcode `\^12\catcode `\_12\catcode `\%12\relax}%
\providecommand \@@startlink[1]{}%
\providecommand \@@endlink[0]{}%
\providecommand \url  [0]{\begingroup\@sanitize@url \@url }%
\providecommand \@url [1]{\endgroup\@href {#1}{\urlprefix }}%
\providecommand \urlprefix  [0]{URL }%
\providecommand \Eprint [0]{\href }%
\providecommand \doibase [0]{https://doi.org/}%
\providecommand \selectlanguage [0]{\@gobble}%
\providecommand \bibinfo  [0]{\@secondoftwo}%
\providecommand \bibfield  [0]{\@secondoftwo}%
\providecommand \translation [1]{[#1]}%
\providecommand \BibitemOpen [0]{}%
\providecommand \bibitemStop [0]{}%
\providecommand \bibitemNoStop [0]{.\EOS\space}%
\providecommand \EOS [0]{\spacefactor3000\relax}%
\providecommand \BibitemShut  [1]{\csname bibitem#1\endcsname}%
\let\auto@bib@innerbib\@empty
%</preamble>
\bibitem [{\citenamefont {Petrov}(2015)}]{Petrov_2015}%
  \BibitemOpen
  \bibfield  {author} {\bibinfo {author} {\bibfnamefont {D.~S.}\ \bibnamefont {Petrov}},\ }\bibfield  {title} {\bibinfo {title} {Quantum mechanical stabilization of a collapsing bose-bose mixture},\ }\bibfield  {journal} {\bibinfo  {journal} {Physical Review Letters}\ }\textbf {\bibinfo {volume} {115}},\ \href {https://doi.org/10.1103/physrevlett.115.155302} {10.1103/physrevlett.115.155302} (\bibinfo {year} {2015})\BibitemShut {NoStop}%
\bibitem [{\citenamefont {Kadau}\ \emph {et~al.}(2016)\citenamefont {Kadau}, \citenamefont {Schmitt}, \citenamefont {Wenzel}, \citenamefont {Wink}, \citenamefont {Maier}, \citenamefont {Ferrier-Barbut},\ and\ \citenamefont {Pfau}}]{Kadau_2016}%
  \BibitemOpen
  \bibfield  {author} {\bibinfo {author} {\bibfnamefont {H.}~\bibnamefont {Kadau}}, \bibinfo {author} {\bibfnamefont {M.}~\bibnamefont {Schmitt}}, \bibinfo {author} {\bibfnamefont {M.}~\bibnamefont {Wenzel}}, \bibinfo {author} {\bibfnamefont {C.}~\bibnamefont {Wink}}, \bibinfo {author} {\bibfnamefont {T.}~\bibnamefont {Maier}}, \bibinfo {author} {\bibfnamefont {I.}~\bibnamefont {Ferrier-Barbut}},\ and\ \bibinfo {author} {\bibfnamefont {T.}~\bibnamefont {Pfau}},\ }\bibfield  {title} {\bibinfo {title} {Observing the rosensweig instability of a quantum ferrofluid},\ }\href {https://doi.org/10.1038/nature16485} {\bibfield  {journal} {\bibinfo  {journal} {Nature}\ }\textbf {\bibinfo {volume} {530}},\ \bibinfo {pages} {194–197} (\bibinfo {year} {2016})}\BibitemShut {NoStop}%
\bibitem [{\citenamefont {Chomaz}\ \emph {et~al.}(2016)\citenamefont {Chomaz}, \citenamefont {Baier}, \citenamefont {Petter}, \citenamefont {Mark}, \citenamefont {Wächtler}, \citenamefont {Santos},\ and\ \citenamefont {Ferlaino}}]{Chomaz_2016}%
  \BibitemOpen
  \bibfield  {author} {\bibinfo {author} {\bibfnamefont {L.}~\bibnamefont {Chomaz}}, \bibinfo {author} {\bibfnamefont {S.}~\bibnamefont {Baier}}, \bibinfo {author} {\bibfnamefont {D.}~\bibnamefont {Petter}}, \bibinfo {author} {\bibfnamefont {M.}~\bibnamefont {Mark}}, \bibinfo {author} {\bibfnamefont {F.}~\bibnamefont {Wächtler}}, \bibinfo {author} {\bibfnamefont {L.}~\bibnamefont {Santos}},\ and\ \bibinfo {author} {\bibfnamefont {F.}~\bibnamefont {Ferlaino}},\ }\bibfield  {title} {\bibinfo {title} {Quantum-fluctuation-driven crossover from a dilute bose-einstein condensate to a macrodroplet in a dipolar quantum fluid},\ }\bibfield  {journal} {\bibinfo  {journal} {Physical Review X}\ }\textbf {\bibinfo {volume} {6}},\ \href {https://doi.org/10.1103/physrevx.6.041039} {10.1103/physrevx.6.041039} (\bibinfo {year} {2016})\BibitemShut {NoStop}%
\bibitem [{\citenamefont {Ferrier-Barbut}\ \emph {et~al.}(2018)\citenamefont {Ferrier-Barbut}, \citenamefont {Wenzel}, \citenamefont {Böttcher}, \citenamefont {Langen}, \citenamefont {Isoard}, \citenamefont {Stringari},\ and\ \citenamefont {Pfau}}]{Ferrier_Barbut_2018}%
  \BibitemOpen
  \bibfield  {author} {\bibinfo {author} {\bibfnamefont {I.}~\bibnamefont {Ferrier-Barbut}}, \bibinfo {author} {\bibfnamefont {M.}~\bibnamefont {Wenzel}}, \bibinfo {author} {\bibfnamefont {F.}~\bibnamefont {Böttcher}}, \bibinfo {author} {\bibfnamefont {T.}~\bibnamefont {Langen}}, \bibinfo {author} {\bibfnamefont {M.}~\bibnamefont {Isoard}}, \bibinfo {author} {\bibfnamefont {S.}~\bibnamefont {Stringari}},\ and\ \bibinfo {author} {\bibfnamefont {T.}~\bibnamefont {Pfau}},\ }\bibfield  {title} {\bibinfo {title} {Scissors mode of dipolar quantum droplets of dysprosium atoms},\ }\bibfield  {journal} {\bibinfo  {journal} {Physical Review Letters}\ }\textbf {\bibinfo {volume} {120}},\ \href {https://doi.org/10.1103/physrevlett.120.160402} {10.1103/physrevlett.120.160402} (\bibinfo {year} {2018})\BibitemShut {NoStop}%
\bibitem [{\citenamefont {Semeghini}\ \emph {et~al.}(2018)\citenamefont {Semeghini}, \citenamefont {Ferioli}, \citenamefont {Masi}, \citenamefont {Mazzinghi}, \citenamefont {Wolswijk}, \citenamefont {Minardi}, \citenamefont {Modugno}, \citenamefont {Modugno}, \citenamefont {Inguscio},\ and\ \citenamefont {Fattori}}]{PhysRevLett.120.235301}%
  \BibitemOpen
  \bibfield  {author} {\bibinfo {author} {\bibfnamefont {G.}~\bibnamefont {Semeghini}}, \bibinfo {author} {\bibfnamefont {G.}~\bibnamefont {Ferioli}}, \bibinfo {author} {\bibfnamefont {L.}~\bibnamefont {Masi}}, \bibinfo {author} {\bibfnamefont {C.}~\bibnamefont {Mazzinghi}}, \bibinfo {author} {\bibfnamefont {L.}~\bibnamefont {Wolswijk}}, \bibinfo {author} {\bibfnamefont {F.}~\bibnamefont {Minardi}}, \bibinfo {author} {\bibfnamefont {M.}~\bibnamefont {Modugno}}, \bibinfo {author} {\bibfnamefont {G.}~\bibnamefont {Modugno}}, \bibinfo {author} {\bibfnamefont {M.}~\bibnamefont {Inguscio}},\ and\ \bibinfo {author} {\bibfnamefont {M.}~\bibnamefont {Fattori}},\ }\bibfield  {title} {\bibinfo {title} {Self-bound quantum droplets of atomic mixtures in free space},\ }\href {https://doi.org/10.1103/PhysRevLett.120.235301} {\bibfield  {journal} {\bibinfo  {journal} {Phys. Rev. Lett.}\ }\textbf {\bibinfo {volume} {120}},\ \bibinfo {pages} {235301} (\bibinfo {year} {2018})}\BibitemShut {NoStop}%
\bibitem [{\citenamefont {Cheiney}\ \emph {et~al.}(2018)\citenamefont {Cheiney}, \citenamefont {Cabrera}, \citenamefont {Sanz}, \citenamefont {Naylor}, \citenamefont {Tanzi},\ and\ \citenamefont {Tarruell}}]{PhysRevLett.120.135301}%
  \BibitemOpen
  \bibfield  {author} {\bibinfo {author} {\bibfnamefont {P.}~\bibnamefont {Cheiney}}, \bibinfo {author} {\bibfnamefont {C.~R.}\ \bibnamefont {Cabrera}}, \bibinfo {author} {\bibfnamefont {J.}~\bibnamefont {Sanz}}, \bibinfo {author} {\bibfnamefont {B.}~\bibnamefont {Naylor}}, \bibinfo {author} {\bibfnamefont {L.}~\bibnamefont {Tanzi}},\ and\ \bibinfo {author} {\bibfnamefont {L.}~\bibnamefont {Tarruell}},\ }\bibfield  {title} {\bibinfo {title} {Bright soliton to quantum droplet transition in a mixture of bose-einstein condensates},\ }\href {https://doi.org/10.1103/PhysRevLett.120.135301} {\bibfield  {journal} {\bibinfo  {journal} {Phys. Rev. Lett.}\ }\textbf {\bibinfo {volume} {120}},\ \bibinfo {pages} {135301} (\bibinfo {year} {2018})}\BibitemShut {NoStop}%
\bibitem [{\citenamefont {Böttcher}\ \emph {et~al.}(2019)\citenamefont {Böttcher}, \citenamefont {Wenzel}, \citenamefont {Schmidt}, \citenamefont {Guo}, \citenamefont {Langen}, \citenamefont {Ferrier-Barbut}, \citenamefont {Pfau}, \citenamefont {Bombín}, \citenamefont {Sánchez-Baena}, \citenamefont {Boronat},\ and\ \citenamefont {Mazzanti}}]{Bottcher_2019}%
  \BibitemOpen
  \bibfield  {author} {\bibinfo {author} {\bibfnamefont {F.}~\bibnamefont {Böttcher}}, \bibinfo {author} {\bibfnamefont {M.}~\bibnamefont {Wenzel}}, \bibinfo {author} {\bibfnamefont {J.-N.}\ \bibnamefont {Schmidt}}, \bibinfo {author} {\bibfnamefont {M.}~\bibnamefont {Guo}}, \bibinfo {author} {\bibfnamefont {T.}~\bibnamefont {Langen}}, \bibinfo {author} {\bibfnamefont {I.}~\bibnamefont {Ferrier-Barbut}}, \bibinfo {author} {\bibfnamefont {T.}~\bibnamefont {Pfau}}, \bibinfo {author} {\bibfnamefont {R.}~\bibnamefont {Bombín}}, \bibinfo {author} {\bibfnamefont {J.}~\bibnamefont {Sánchez-Baena}}, \bibinfo {author} {\bibfnamefont {J.}~\bibnamefont {Boronat}},\ and\ \bibinfo {author} {\bibfnamefont {F.}~\bibnamefont {Mazzanti}},\ }\bibfield  {title} {\bibinfo {title} {Dilute dipolar quantum droplets beyond the extended gross-pitaevskii equation},\ }\bibfield  {journal} {\bibinfo  {journal} {Physical Review Research}\ }\textbf {\bibinfo {volume} {1}},\ \href {https://doi.org/10.1103/physrevresearch.1.033088}
  {10.1103/physrevresearch.1.033088} (\bibinfo {year} {2019})\BibitemShut {NoStop}%
\bibitem [{\citenamefont {W\"achtler}\ and\ \citenamefont {Santos}(2016)}]{PhysRevA.93.061603}%
  \BibitemOpen
  \bibfield  {author} {\bibinfo {author} {\bibfnamefont {F.}~\bibnamefont {W\"achtler}}\ and\ \bibinfo {author} {\bibfnamefont {L.}~\bibnamefont {Santos}},\ }\bibfield  {title} {\bibinfo {title} {Quantum filaments in dipolar bose-einstein condensates},\ }\href {https://doi.org/10.1103/PhysRevA.93.061603} {\bibfield  {journal} {\bibinfo  {journal} {Phys. Rev. A}\ }\textbf {\bibinfo {volume} {93}},\ \bibinfo {pages} {061603} (\bibinfo {year} {2016})}\BibitemShut {NoStop}%
\bibitem [{\citenamefont {Baillie}\ \emph {et~al.}(2016)\citenamefont {Baillie}, \citenamefont {Wilson}, \citenamefont {Bisset},\ and\ \citenamefont {Blakie}}]{PhysRevA.94.021602}%
  \BibitemOpen
  \bibfield  {author} {\bibinfo {author} {\bibfnamefont {D.}~\bibnamefont {Baillie}}, \bibinfo {author} {\bibfnamefont {R.~M.}\ \bibnamefont {Wilson}}, \bibinfo {author} {\bibfnamefont {R.~N.}\ \bibnamefont {Bisset}},\ and\ \bibinfo {author} {\bibfnamefont {P.~B.}\ \bibnamefont {Blakie}},\ }\bibfield  {title} {\bibinfo {title} {Self-bound dipolar droplet: A localized matter wave in free space},\ }\href {https://doi.org/10.1103/PhysRevA.94.021602} {\bibfield  {journal} {\bibinfo  {journal} {Phys. Rev. A}\ }\textbf {\bibinfo {volume} {94}},\ \bibinfo {pages} {021602} (\bibinfo {year} {2016})}\BibitemShut {NoStop}%
\bibitem [{\citenamefont {Baillie}\ \emph {et~al.}(2017)\citenamefont {Baillie}, \citenamefont {Wilson},\ and\ \citenamefont {Blakie}}]{PhysRevLett.119.255302}%
  \BibitemOpen
  \bibfield  {author} {\bibinfo {author} {\bibfnamefont {D.}~\bibnamefont {Baillie}}, \bibinfo {author} {\bibfnamefont {R.~M.}\ \bibnamefont {Wilson}},\ and\ \bibinfo {author} {\bibfnamefont {P.~B.}\ \bibnamefont {Blakie}},\ }\bibfield  {title} {\bibinfo {title} {Collective excitations of self-bound droplets of a dipolar quantum fluid},\ }\href {https://doi.org/10.1103/PhysRevLett.119.255302} {\bibfield  {journal} {\bibinfo  {journal} {Phys. Rev. Lett.}\ }\textbf {\bibinfo {volume} {119}},\ \bibinfo {pages} {255302} (\bibinfo {year} {2017})}\BibitemShut {NoStop}%
\bibitem [{\citenamefont {Lee}\ \emph {et~al.}(1957)\citenamefont {Lee}, \citenamefont {Huang},\ and\ \citenamefont {Yang}}]{lhy}%
  \BibitemOpen
  \bibfield  {author} {\bibinfo {author} {\bibfnamefont {T.~D.}\ \bibnamefont {Lee}}, \bibinfo {author} {\bibfnamefont {K.}~\bibnamefont {Huang}},\ and\ \bibinfo {author} {\bibfnamefont {C.~N.}\ \bibnamefont {Yang}},\ }\bibfield  {title} {\bibinfo {title} {Eigenvalues and eigenfunctions of a bose system of hard spheres and its low-temperature properties},\ }\href {https://doi.org/10.1103/PhysRev.106.1135} {\bibfield  {journal} {\bibinfo  {journal} {Phys. Rev.}\ }\textbf {\bibinfo {volume} {106}},\ \bibinfo {pages} {1135} (\bibinfo {year} {1957})}\BibitemShut {NoStop}%
\bibitem [{\citenamefont {Schützhold}\ \emph {et~al.}(2006)\citenamefont {Schützhold}, \citenamefont {Uhlmann}, \citenamefont {Xu},\ and\ \citenamefont {Fischer}}]{SCH_TZHOLD_2006}%
  \BibitemOpen
  \bibfield  {author} {\bibinfo {author} {\bibfnamefont {R.}~\bibnamefont {Schützhold}}, \bibinfo {author} {\bibfnamefont {M.}~\bibnamefont {Uhlmann}}, \bibinfo {author} {\bibfnamefont {Y.}~\bibnamefont {Xu}},\ and\ \bibinfo {author} {\bibfnamefont {U.~R.}\ \bibnamefont {Fischer}},\ }\bibfield  {title} {\bibinfo {title} {Mean-field expansion in bose-einstein condensates with finite-range interactions},\ }\href {https://doi.org/10.1142/s0217979206035631} {\bibfield  {journal} {\bibinfo  {journal} {International Journal of Modern Physics B}\ }\textbf {\bibinfo {volume} {20}},\ \bibinfo {pages} {3555–3565} (\bibinfo {year} {2006})}\BibitemShut {NoStop}%
\bibitem [{\citenamefont {Lima}\ and\ \citenamefont {Pelster}(2011)}]{Lima_2011}%
  \BibitemOpen
  \bibfield  {author} {\bibinfo {author} {\bibfnamefont {A.~R.~P.}\ \bibnamefont {Lima}}\ and\ \bibinfo {author} {\bibfnamefont {A.}~\bibnamefont {Pelster}},\ }\bibfield  {title} {\bibinfo {title} {Quantum fluctuations in dipolar bose gases},\ }\bibfield  {journal} {\bibinfo  {journal} {Physical Review A}\ }\textbf {\bibinfo {volume} {84}},\ \href {https://doi.org/10.1103/physreva.84.041604} {10.1103/physreva.84.041604} (\bibinfo {year} {2011})\BibitemShut {NoStop}%
\bibitem [{\citenamefont {Lima}\ and\ \citenamefont {Pelster}(2012)}]{PhysRevA.86.063609}%
  \BibitemOpen
  \bibfield  {author} {\bibinfo {author} {\bibfnamefont {A.~R.~P.}\ \bibnamefont {Lima}}\ and\ \bibinfo {author} {\bibfnamefont {A.}~\bibnamefont {Pelster}},\ }\bibfield  {title} {\bibinfo {title} {Beyond mean-field low-lying excitations of dipolar bose gases},\ }\href {https://doi.org/10.1103/PhysRevA.86.063609} {\bibfield  {journal} {\bibinfo  {journal} {Phys. Rev. A}\ }\textbf {\bibinfo {volume} {86}},\ \bibinfo {pages} {063609} (\bibinfo {year} {2012})}\BibitemShut {NoStop}%
\bibitem [{\citenamefont {Naraschewski}\ and\ \citenamefont {Glauber}(1999)}]{Naraschewski_1999}%
  \BibitemOpen
  \bibfield  {author} {\bibinfo {author} {\bibfnamefont {M.}~\bibnamefont {Naraschewski}}\ and\ \bibinfo {author} {\bibfnamefont {R.~J.}\ \bibnamefont {Glauber}},\ }\bibfield  {title} {\bibinfo {title} {Spatial coherence and density correlations of trapped bose gases},\ }\href {https://doi.org/10.1103/physreva.59.4595} {\bibfield  {journal} {\bibinfo  {journal} {Physical Review A}\ }\textbf {\bibinfo {volume} {59}},\ \bibinfo {pages} {4595–4607} (\bibinfo {year} {1999})}\BibitemShut {NoStop}%
\bibitem [{\citenamefont {Manz}\ \emph {et~al.}(2010)\citenamefont {Manz}, \citenamefont {B\"ucker}, \citenamefont {Betz}, \citenamefont {Koller}, \citenamefont {Hofferberth}, \citenamefont {Mazets}, \citenamefont {Imambekov}, \citenamefont {Demler}, \citenamefont {Perrin}, \citenamefont {Schmiedmayer},\ and\ \citenamefont {Schumm}}]{PhysRevA.81.031610}%
  \BibitemOpen
  \bibfield  {author} {\bibinfo {author} {\bibfnamefont {S.}~\bibnamefont {Manz}}, \bibinfo {author} {\bibfnamefont {R.}~\bibnamefont {B\"ucker}}, \bibinfo {author} {\bibfnamefont {T.}~\bibnamefont {Betz}}, \bibinfo {author} {\bibfnamefont {C.}~\bibnamefont {Koller}}, \bibinfo {author} {\bibfnamefont {S.}~\bibnamefont {Hofferberth}}, \bibinfo {author} {\bibfnamefont {I.~E.}\ \bibnamefont {Mazets}}, \bibinfo {author} {\bibfnamefont {A.}~\bibnamefont {Imambekov}}, \bibinfo {author} {\bibfnamefont {E.}~\bibnamefont {Demler}}, \bibinfo {author} {\bibfnamefont {A.}~\bibnamefont {Perrin}}, \bibinfo {author} {\bibfnamefont {J.}~\bibnamefont {Schmiedmayer}},\ and\ \bibinfo {author} {\bibfnamefont {T.}~\bibnamefont {Schumm}},\ }\bibfield  {title} {\bibinfo {title} {Two-point density correlations of quasicondensates in free expansion},\ }\href {https://doi.org/10.1103/PhysRevA.81.031610} {\bibfield  {journal} {\bibinfo  {journal} {Phys. Rev. A}\ }\textbf {\bibinfo {volume} {81}},\ \bibinfo {pages} {031610} (\bibinfo
  {year} {2010})}\BibitemShut {NoStop}%
\bibitem [{\citenamefont {Holas}\ \emph {et~al.}(1979)\citenamefont {Holas}, \citenamefont {Aravind},\ and\ \citenamefont {Singwi}}]{PhysRevB.20.4912}%
  \BibitemOpen
  \bibfield  {author} {\bibinfo {author} {\bibfnamefont {A.}~\bibnamefont {Holas}}, \bibinfo {author} {\bibfnamefont {P.~K.}\ \bibnamefont {Aravind}},\ and\ \bibinfo {author} {\bibfnamefont {K.~S.}\ \bibnamefont {Singwi}},\ }\bibfield  {title} {\bibinfo {title} {Dynamic correlations in an electron gas. i. first-order perturbation theory},\ }\href {https://doi.org/10.1103/PhysRevB.20.4912} {\bibfield  {journal} {\bibinfo  {journal} {Phys. Rev. B}\ }\textbf {\bibinfo {volume} {20}},\ \bibinfo {pages} {4912} (\bibinfo {year} {1979})}\BibitemShut {NoStop}%
\bibitem [{\citenamefont {Hung}\ \emph {et~al.}(2011)\citenamefont {Hung}, \citenamefont {Zhang}, \citenamefont {Ha}, \citenamefont {Tung}, \citenamefont {Gemelke},\ and\ \citenamefont {Chin}}]{Hung_2011}%
  \BibitemOpen
  \bibfield  {author} {\bibinfo {author} {\bibfnamefont {C.-L.}\ \bibnamefont {Hung}}, \bibinfo {author} {\bibfnamefont {X.}~\bibnamefont {Zhang}}, \bibinfo {author} {\bibfnamefont {L.-C.}\ \bibnamefont {Ha}}, \bibinfo {author} {\bibfnamefont {S.-K.}\ \bibnamefont {Tung}}, \bibinfo {author} {\bibfnamefont {N.}~\bibnamefont {Gemelke}},\ and\ \bibinfo {author} {\bibfnamefont {C.}~\bibnamefont {Chin}},\ }\bibfield  {title} {\bibinfo {title} {Extracting density–density correlations from in situ images of atomic quantum gases},\ }\href {https://doi.org/10.1088/1367-2630/13/7/075019} {\bibfield  {journal} {\bibinfo  {journal} {New Journal of Physics}\ }\textbf {\bibinfo {volume} {13}},\ \bibinfo {pages} {075019} (\bibinfo {year} {2011})}\BibitemShut {NoStop}%
\bibitem [{\citenamefont {Dowling}\ \emph {et~al.}(2006)\citenamefont {Dowling}, \citenamefont {Doherty},\ and\ \citenamefont {Wiseman}}]{Dowling_2006}%
  \BibitemOpen
  \bibfield  {author} {\bibinfo {author} {\bibfnamefont {M.~R.}\ \bibnamefont {Dowling}}, \bibinfo {author} {\bibfnamefont {A.~C.}\ \bibnamefont {Doherty}},\ and\ \bibinfo {author} {\bibfnamefont {H.~M.}\ \bibnamefont {Wiseman}},\ }\bibfield  {title} {\bibinfo {title} {Entanglement of indistinguishable particles in condensed-matter physics},\ }\bibfield  {journal} {\bibinfo  {journal} {Physical Review A}\ }\textbf {\bibinfo {volume} {73}},\ \href {https://doi.org/10.1103/physreva.73.052323} {10.1103/physreva.73.052323} (\bibinfo {year} {2006})\BibitemShut {NoStop}%
\bibitem [{\citenamefont {Amico}\ \emph {et~al.}(2008)\citenamefont {Amico}, \citenamefont {Fazio}, \citenamefont {Osterloh},\ and\ \citenamefont {Vedral}}]{RevModPhys.80.517}%
  \BibitemOpen
  \bibfield  {author} {\bibinfo {author} {\bibfnamefont {L.}~\bibnamefont {Amico}}, \bibinfo {author} {\bibfnamefont {R.}~\bibnamefont {Fazio}}, \bibinfo {author} {\bibfnamefont {A.}~\bibnamefont {Osterloh}},\ and\ \bibinfo {author} {\bibfnamefont {V.}~\bibnamefont {Vedral}},\ }\bibfield  {title} {\bibinfo {title} {Entanglement in many-body systems},\ }\href {https://doi.org/10.1103/RevModPhys.80.517} {\bibfield  {journal} {\bibinfo  {journal} {Rev. Mod. Phys.}\ }\textbf {\bibinfo {volume} {80}},\ \bibinfo {pages} {517} (\bibinfo {year} {2008})}\BibitemShut {NoStop}%
\bibitem [{\citenamefont {Norcia}\ \emph {et~al.}(2021)\citenamefont {Norcia}, \citenamefont {Politi}, \citenamefont {Klaus}, \citenamefont {Poli}, \citenamefont {Sohmen}, \citenamefont {Mark}, \citenamefont {Bisset}, \citenamefont {Santos},\ and\ \citenamefont {Ferlaino}}]{Norcia_2021}%
  \BibitemOpen
  \bibfield  {author} {\bibinfo {author} {\bibfnamefont {M.~A.}\ \bibnamefont {Norcia}}, \bibinfo {author} {\bibfnamefont {C.}~\bibnamefont {Politi}}, \bibinfo {author} {\bibfnamefont {L.}~\bibnamefont {Klaus}}, \bibinfo {author} {\bibfnamefont {E.}~\bibnamefont {Poli}}, \bibinfo {author} {\bibfnamefont {M.}~\bibnamefont {Sohmen}}, \bibinfo {author} {\bibfnamefont {M.~J.}\ \bibnamefont {Mark}}, \bibinfo {author} {\bibfnamefont {R.~N.}\ \bibnamefont {Bisset}}, \bibinfo {author} {\bibfnamefont {L.}~\bibnamefont {Santos}},\ and\ \bibinfo {author} {\bibfnamefont {F.}~\bibnamefont {Ferlaino}},\ }\bibfield  {title} {\bibinfo {title} {Two-dimensional supersolidity in a dipolar quantum gas},\ }\href {https://doi.org/10.1038/s41586-021-03725-7} {\bibfield  {journal} {\bibinfo  {journal} {Nature}\ }\textbf {\bibinfo {volume} {596}},\ \bibinfo {pages} {357} (\bibinfo {year} {2021})}\BibitemShut {NoStop}%
\bibitem [{\citenamefont {Bland}\ \emph {et~al.}(2022)\citenamefont {Bland}, \citenamefont {Poli}, \citenamefont {Politi}, \citenamefont {Klaus}, \citenamefont {Norcia}, \citenamefont {Ferlaino}, \citenamefont {Santos},\ and\ \citenamefont {Bisset}}]{Bland_2022}%
  \BibitemOpen
  \bibfield  {author} {\bibinfo {author} {\bibfnamefont {T.}~\bibnamefont {Bland}}, \bibinfo {author} {\bibfnamefont {E.}~\bibnamefont {Poli}}, \bibinfo {author} {\bibfnamefont {C.}~\bibnamefont {Politi}}, \bibinfo {author} {\bibfnamefont {L.}~\bibnamefont {Klaus}}, \bibinfo {author} {\bibfnamefont {M.}~\bibnamefont {Norcia}}, \bibinfo {author} {\bibfnamefont {F.}~\bibnamefont {Ferlaino}}, \bibinfo {author} {\bibfnamefont {L.}~\bibnamefont {Santos}},\ and\ \bibinfo {author} {\bibfnamefont {R.}~\bibnamefont {Bisset}},\ }\bibfield  {title} {\bibinfo {title} {Two-dimensional supersolid formation in dipolar condensates},\ }\bibfield  {journal} {\bibinfo  {journal} {Physical Review Letters}\ }\textbf {\bibinfo {volume} {128}},\ \href {https://doi.org/10.1103/physrevlett.128.195302} {10.1103/physrevlett.128.195302} (\bibinfo {year} {2022})\BibitemShut {NoStop}%
\bibitem [{\citenamefont {Schmitt}\ \emph {et~al.}(2016)\citenamefont {Schmitt}, \citenamefont {Wenzel}, \citenamefont {Böttcher}, \citenamefont {Ferrier-Barbut},\ and\ \citenamefont {Pfau}}]{Schmitt_2016}%
  \BibitemOpen
  \bibfield  {author} {\bibinfo {author} {\bibfnamefont {M.}~\bibnamefont {Schmitt}}, \bibinfo {author} {\bibfnamefont {M.}~\bibnamefont {Wenzel}}, \bibinfo {author} {\bibfnamefont {F.}~\bibnamefont {Böttcher}}, \bibinfo {author} {\bibfnamefont {I.}~\bibnamefont {Ferrier-Barbut}},\ and\ \bibinfo {author} {\bibfnamefont {T.}~\bibnamefont {Pfau}},\ }\bibfield  {title} {\bibinfo {title} {Self-bound droplets of a dilute magnetic quantum liquid},\ }\href {https://doi.org/10.1038/nature20126} {\bibfield  {journal} {\bibinfo  {journal} {Nature}\ }\textbf {\bibinfo {volume} {539}},\ \bibinfo {pages} {259–262} (\bibinfo {year} {2016})}\BibitemShut {NoStop}%
\bibitem [{\citenamefont {Tanzi}\ \emph {et~al.}(2019)\citenamefont {Tanzi}, \citenamefont {Lucioni}, \citenamefont {Fam\`a}, \citenamefont {Catani}, \citenamefont {Fioretti}, \citenamefont {Gabbanini}, \citenamefont {Bisset}, \citenamefont {Santos},\ and\ \citenamefont {Modugno}}]{Modugno_2019}%
  \BibitemOpen
  \bibfield  {author} {\bibinfo {author} {\bibfnamefont {L.}~\bibnamefont {Tanzi}}, \bibinfo {author} {\bibfnamefont {E.}~\bibnamefont {Lucioni}}, \bibinfo {author} {\bibfnamefont {F.}~\bibnamefont {Fam\`a}}, \bibinfo {author} {\bibfnamefont {J.}~\bibnamefont {Catani}}, \bibinfo {author} {\bibfnamefont {A.}~\bibnamefont {Fioretti}}, \bibinfo {author} {\bibfnamefont {C.}~\bibnamefont {Gabbanini}}, \bibinfo {author} {\bibfnamefont {R.~N.}\ \bibnamefont {Bisset}}, \bibinfo {author} {\bibfnamefont {L.}~\bibnamefont {Santos}},\ and\ \bibinfo {author} {\bibfnamefont {G.}~\bibnamefont {Modugno}},\ }\bibfield  {title} {\bibinfo {title} {Observation of a dipolar quantum gas with metastable supersolid properties},\ }\href {https://doi.org/10.1103/PhysRevLett.122.130405} {\bibfield  {journal} {\bibinfo  {journal} {Phys. Rev. Lett.}\ }\textbf {\bibinfo {volume} {122}},\ \bibinfo {pages} {130405} (\bibinfo {year} {2019})}\BibitemShut {NoStop}%
\bibitem [{\citenamefont {Petrov}\ and\ \citenamefont {Astrakharchik}(2016)}]{Petrov2016}%
  \BibitemOpen
  \bibfield  {author} {\bibinfo {author} {\bibfnamefont {D.~S.}\ \bibnamefont {Petrov}}\ and\ \bibinfo {author} {\bibfnamefont {G.~E.}\ \bibnamefont {Astrakharchik}},\ }\bibfield  {title} {\bibinfo {title} {Ultradilute low-dimensional liquids},\ }\href {https://doi.org/10.1103/PhysRevLett.117.100401} {\bibfield  {journal} {\bibinfo  {journal} {Phys. Rev. Lett.}\ }\textbf {\bibinfo {volume} {117}},\ \bibinfo {pages} {100401} (\bibinfo {year} {2016})}\BibitemShut {NoStop}%
\bibitem [{\citenamefont {Parisi}\ \emph {et~al.}(2019)\citenamefont {Parisi}, \citenamefont {Astrakharchik},\ and\ \citenamefont {Giorgini}}]{Parisi2019}%
  \BibitemOpen
  \bibfield  {author} {\bibinfo {author} {\bibfnamefont {L.}~\bibnamefont {Parisi}}, \bibinfo {author} {\bibfnamefont {G.~E.}\ \bibnamefont {Astrakharchik}},\ and\ \bibinfo {author} {\bibfnamefont {S.}~\bibnamefont {Giorgini}},\ }\bibfield  {title} {\bibinfo {title} {Liquid state of one-dimensional bose mixtures: A quantum monte carlo study},\ }\href {https://doi.org/10.1103/PhysRevLett.122.105302} {\bibfield  {journal} {\bibinfo  {journal} {Phys. Rev. Lett.}\ }\textbf {\bibinfo {volume} {122}},\ \bibinfo {pages} {105302} (\bibinfo {year} {2019})}\BibitemShut {NoStop}%
\bibitem [{\citenamefont {Parisi}\ and\ \citenamefont {Giorgini}(2020)}]{Parisi2020}%
  \BibitemOpen
  \bibfield  {author} {\bibinfo {author} {\bibfnamefont {L.}~\bibnamefont {Parisi}}\ and\ \bibinfo {author} {\bibfnamefont {S.}~\bibnamefont {Giorgini}},\ }\bibfield  {title} {\bibinfo {title} {Quantum droplets in one-dimensional bose mixtures: A quantum monte carlo study},\ }\href {https://doi.org/10.1103/PhysRevA.102.023318} {\bibfield  {journal} {\bibinfo  {journal} {Phys. Rev. A}\ }\textbf {\bibinfo {volume} {102}},\ \bibinfo {pages} {023318} (\bibinfo {year} {2020})}\BibitemShut {NoStop}%
\bibitem [{\citenamefont {Kopyci\ifmmode~\acute{n}\else \'{n}\fi{}ski}\ \emph {et~al.}(2023)\citenamefont {Kopyci\ifmmode~\acute{n}\else \'{n}\fi{}ski}, \citenamefont {Parisi}, \citenamefont {Parker},\ and\ \citenamefont {Paw\l{}owski}}]{Kopycinski2023}%
  \BibitemOpen
  \bibfield  {author} {\bibinfo {author} {\bibfnamefont {J.}~\bibnamefont {Kopyci\ifmmode~\acute{n}\else \'{n}\fi{}ski}}, \bibinfo {author} {\bibfnamefont {L.}~\bibnamefont {Parisi}}, \bibinfo {author} {\bibfnamefont {N.~G.}\ \bibnamefont {Parker}},\ and\ \bibinfo {author} {\bibfnamefont {K.}~\bibnamefont {Paw\l{}owski}},\ }\bibfield  {title} {\bibinfo {title} {Quantum monte carlo-based density functional for one-dimensional bose-bose mixtures},\ }\href {https://doi.org/10.1103/PhysRevResearch.5.023050} {\bibfield  {journal} {\bibinfo  {journal} {Phys. Rev. Res.}\ }\textbf {\bibinfo {volume} {5}},\ \bibinfo {pages} {023050} (\bibinfo {year} {2023})}\BibitemShut {NoStop}%
\bibitem [{\citenamefont {O{\l}dziejewski}\ \emph {et~al.}(2020)\citenamefont {O{\l}dziejewski}, \citenamefont {G{\'{o}}recki}, \citenamefont {Paw{\l}owski},\ and\ \citenamefont {Rz{\k{a}}{\.{z}}ewski}}]{Oldziejewski_2020}%
  \BibitemOpen
  \bibfield  {author} {\bibinfo {author} {\bibfnamefont {R.}~\bibnamefont {O{\l}dziejewski}}, \bibinfo {author} {\bibfnamefont {W.}~\bibnamefont {G{\'{o}}recki}}, \bibinfo {author} {\bibfnamefont {K.}~\bibnamefont {Paw{\l}owski}},\ and\ \bibinfo {author} {\bibfnamefont {K.}~\bibnamefont {Rz{\k{a}}{\.{z}}ewski}},\ }\bibfield  {title} {\bibinfo {title} {Strongly correlated quantum droplets in quasi-1d dipolar bose gas},\ }\bibfield  {journal} {\bibinfo  {journal} {Physical Review Letters}\ }\textbf {\bibinfo {volume} {124}},\ \href {https://doi.org/10.1103/physrevlett.124.090401} {10.1103/physrevlett.124.090401} (\bibinfo {year} {2020})\BibitemShut {NoStop}%
\bibitem [{\citenamefont {Baizakov}\ \emph {et~al.}(2009)\citenamefont {Baizakov}, \citenamefont {Abdullaev}, \citenamefont {Malomed},\ and\ \citenamefont {Salerno}}]{baizakov_solitons_2009}%
  \BibitemOpen
  \bibfield  {author} {\bibinfo {author} {\bibfnamefont {B.~B.}\ \bibnamefont {Baizakov}}, \bibinfo {author} {\bibfnamefont {F.~K.}\ \bibnamefont {Abdullaev}}, \bibinfo {author} {\bibfnamefont {B.~A.}\ \bibnamefont {Malomed}},\ and\ \bibinfo {author} {\bibfnamefont {M.}~\bibnamefont {Salerno}},\ }\bibfield  {title} {\bibinfo {title} {Solitons in the tonks–girardeau gas with dipolar interactions},\ }\href {https://doi.org/10.1088/0953-4075/42/17/175302} {\bibfield  {journal} {\bibinfo  {journal} {Journal of Physics B: Atomic, Molecular and Optical Physics}\ }\textbf {\bibinfo {volume} {42}},\ \bibinfo {pages} {175302} (\bibinfo {year} {2009})}\BibitemShut {NoStop}%
\bibitem [{\citenamefont {De~Palo}\ \emph {et~al.}(2020)\citenamefont {De~Palo}, \citenamefont {Citro},\ and\ \citenamefont {Orignac}}]{dePalo2020}%
  \BibitemOpen
  \bibfield  {author} {\bibinfo {author} {\bibfnamefont {S.}~\bibnamefont {De~Palo}}, \bibinfo {author} {\bibfnamefont {R.}~\bibnamefont {Citro}},\ and\ \bibinfo {author} {\bibfnamefont {E.}~\bibnamefont {Orignac}},\ }\bibfield  {title} {\bibinfo {title} {Variational bethe ansatz approach for dipolar one-dimensional bosons},\ }\href {https://doi.org/10.1103/PhysRevB.101.045102} {\bibfield  {journal} {\bibinfo  {journal} {Phys. Rev. B}\ }\textbf {\bibinfo {volume} {101}},\ \bibinfo {pages} {045102} (\bibinfo {year} {2020})}\BibitemShut {NoStop}%
\bibitem [{\citenamefont {Edler}\ \emph {et~al.}(2017)\citenamefont {Edler}, \citenamefont {Mishra}, \citenamefont {Wächtler}, \citenamefont {Nath}, \citenamefont {Sinha},\ and\ \citenamefont {Santos}}]{Edler_2017}%
  \BibitemOpen
  \bibfield  {author} {\bibinfo {author} {\bibfnamefont {D.}~\bibnamefont {Edler}}, \bibinfo {author} {\bibfnamefont {C.}~\bibnamefont {Mishra}}, \bibinfo {author} {\bibfnamefont {F.}~\bibnamefont {Wächtler}}, \bibinfo {author} {\bibfnamefont {R.}~\bibnamefont {Nath}}, \bibinfo {author} {\bibfnamefont {S.}~\bibnamefont {Sinha}},\ and\ \bibinfo {author} {\bibfnamefont {L.}~\bibnamefont {Santos}},\ }\bibfield  {title} {\bibinfo {title} {Quantum fluctuations in quasi-one-dimensional dipolar bose-einstein condensates},\ }\bibfield  {journal} {\bibinfo  {journal} {Physical Review Letters}\ }\textbf {\bibinfo {volume} {119}},\ \href {https://doi.org/10.1103/physrevlett.119.050403} {10.1103/physrevlett.119.050403} (\bibinfo {year} {2017})\BibitemShut {NoStop}%
\bibitem [{\citenamefont {Kao}\ \emph {et~al.}(2021)\citenamefont {Kao}, \citenamefont {Li}, \citenamefont {Lin}, \citenamefont {Gopalakrishnan},\ and\ \citenamefont {Lev}}]{Kao2021Jan}%
  \BibitemOpen
  \bibfield  {author} {\bibinfo {author} {\bibfnamefont {W.}~\bibnamefont {Kao}}, \bibinfo {author} {\bibfnamefont {K.-Y.}\ \bibnamefont {Li}}, \bibinfo {author} {\bibfnamefont {K.-Y.}\ \bibnamefont {Lin}}, \bibinfo {author} {\bibfnamefont {S.}~\bibnamefont {Gopalakrishnan}},\ and\ \bibinfo {author} {\bibfnamefont {B.~L.}\ \bibnamefont {Lev}},\ }\bibfield  {title} {\bibinfo {title} {{Topological pumping of a 1D dipolar gas into strongly correlated prethermal states}},\ }\href {https://doi.org/10.1126/science.abb4928} {\bibfield  {journal} {\bibinfo  {journal} {Science}\ }\textbf {\bibinfo {volume} {371}},\ \bibinfo {pages} {296} (\bibinfo {year} {2021})}\BibitemShut {NoStop}%
\bibitem [{\citenamefont {Li}\ \emph {et~al.}(2023{\natexlab{a}})\citenamefont {Li}, \citenamefont {Zhang}, \citenamefont {Yang}, \citenamefont {Lin}, \citenamefont {Gopalakrishnan}, \citenamefont {Rigol},\ and\ \citenamefont {Lev}}]{kuanYu2023}%
  \BibitemOpen
  \bibfield  {author} {\bibinfo {author} {\bibfnamefont {K.-Y.}\ \bibnamefont {Li}}, \bibinfo {author} {\bibfnamefont {Y.}~\bibnamefont {Zhang}}, \bibinfo {author} {\bibfnamefont {K.}~\bibnamefont {Yang}}, \bibinfo {author} {\bibfnamefont {K.-Y.}\ \bibnamefont {Lin}}, \bibinfo {author} {\bibfnamefont {S.}~\bibnamefont {Gopalakrishnan}}, \bibinfo {author} {\bibfnamefont {M.}~\bibnamefont {Rigol}},\ and\ \bibinfo {author} {\bibfnamefont {B.~L.}\ \bibnamefont {Lev}},\ }\bibfield  {title} {\bibinfo {title} {Rapidity and momentum distributions of one-dimensional dipolar quantum gases},\ }\href {https://doi.org/10.1103/PhysRevA.107.L061302} {\bibfield  {journal} {\bibinfo  {journal} {Phys. Rev. A}\ }\textbf {\bibinfo {volume} {107}},\ \bibinfo {pages} {L061302} (\bibinfo {year} {2023}{\natexlab{a}})}\BibitemShut {NoStop}%
\bibitem [{\citenamefont {Bigagli}\ \emph {et~al.}(2024)\citenamefont {Bigagli}, \citenamefont {Yuan}, \citenamefont {Zhang}, \citenamefont {Bulatovic}, \citenamefont {Karman}, \citenamefont {Stevenson},\ and\ \citenamefont {Will}}]{Bigagli_2024}%
  \BibitemOpen
  \bibfield  {author} {\bibinfo {author} {\bibfnamefont {N.}~\bibnamefont {Bigagli}}, \bibinfo {author} {\bibfnamefont {W.}~\bibnamefont {Yuan}}, \bibinfo {author} {\bibfnamefont {S.}~\bibnamefont {Zhang}}, \bibinfo {author} {\bibfnamefont {B.}~\bibnamefont {Bulatovic}}, \bibinfo {author} {\bibfnamefont {T.}~\bibnamefont {Karman}}, \bibinfo {author} {\bibfnamefont {I.}~\bibnamefont {Stevenson}},\ and\ \bibinfo {author} {\bibfnamefont {S.}~\bibnamefont {Will}},\ }\bibfield  {title} {\bibinfo {title} {Observation of bose–einstein condensation of dipolar molecules},\ }\bibfield  {journal} {\bibinfo  {journal} {Nature}\ }\href {https://doi.org/10.1038/s41586-024-07492-z} {10.1038/s41586-024-07492-z} (\bibinfo {year} {2024})\BibitemShut {NoStop}%
\bibitem [{\citenamefont {Lieb}\ and\ \citenamefont {Liniger}(1963)}]{lieb}%
  \BibitemOpen
  \bibfield  {author} {\bibinfo {author} {\bibfnamefont {E.~H.}\ \bibnamefont {Lieb}}\ and\ \bibinfo {author} {\bibfnamefont {W.}~\bibnamefont {Liniger}},\ }\bibfield  {title} {\bibinfo {title} {Exact analysis of an interacting bose gas. i. the general solution and the ground state},\ }\href {https://doi.org/10.1103/PhysRev.130.1605} {\bibfield  {journal} {\bibinfo  {journal} {Phys. Rev.}\ }\textbf {\bibinfo {volume} {130}},\ \bibinfo {pages} {1605} (\bibinfo {year} {1963})}\BibitemShut {NoStop}%
\bibitem [{\citenamefont {Lieb}(1963)}]{lieb2}%
  \BibitemOpen
  \bibfield  {author} {\bibinfo {author} {\bibfnamefont {E.~H.}\ \bibnamefont {Lieb}},\ }\bibfield  {title} {\bibinfo {title} {Exact analysis of an interacting bose gas. ii. the excitation spectrum},\ }\href {https://doi.org/10.1103/PhysRev.130.1616} {\bibfield  {journal} {\bibinfo  {journal} {Phys. Rev.}\ }\textbf {\bibinfo {volume} {130}},\ \bibinfo {pages} {1616} (\bibinfo {year} {1963})}\BibitemShut {NoStop}%
\bibitem [{\citenamefont {Girardeau}(1960)}]{Girardeau1960}%
  \BibitemOpen
  \bibfield  {author} {\bibinfo {author} {\bibfnamefont {M.}~\bibnamefont {Girardeau}},\ }\bibfield  {title} {\bibinfo {title} {Relationship between systems of impenetrable bosons and fermions in one dimension},\ }\href@noop {} {\bibfield  {journal} {\bibinfo  {journal} {Journal of Mathematical Physics}\ }\textbf {\bibinfo {volume} {1}},\ \bibinfo {pages} {516} (\bibinfo {year} {1960})}\BibitemShut {NoStop}%
\bibitem [{\citenamefont {Deuretzbacher}\ \emph {et~al.}(2010)\citenamefont {Deuretzbacher}, \citenamefont {Cremon},\ and\ \citenamefont {Reimann}}]{PhysRevA.81.063616}%
  \BibitemOpen
  \bibfield  {author} {\bibinfo {author} {\bibfnamefont {F.}~\bibnamefont {Deuretzbacher}}, \bibinfo {author} {\bibfnamefont {J.~C.}\ \bibnamefont {Cremon}},\ and\ \bibinfo {author} {\bibfnamefont {S.~M.}\ \bibnamefont {Reimann}},\ }\bibfield  {title} {\bibinfo {title} {Ground-state properties of few dipolar bosons in a quasi-one-dimensional harmonic trap},\ }\href {https://doi.org/10.1103/PhysRevA.81.063616} {\bibfield  {journal} {\bibinfo  {journal} {Phys. Rev. A}\ }\textbf {\bibinfo {volume} {81}},\ \bibinfo {pages} {063616} (\bibinfo {year} {2010})}\BibitemShut {NoStop}%
\bibitem [{\citenamefont {Deuretzbacher}\ \emph {et~al.}(2013)\citenamefont {Deuretzbacher}, \citenamefont {Cremon},\ and\ \citenamefont {Reimann}}]{PhysRevA.87.039903}%
  \BibitemOpen
  \bibfield  {author} {\bibinfo {author} {\bibfnamefont {F.}~\bibnamefont {Deuretzbacher}}, \bibinfo {author} {\bibfnamefont {J.~C.}\ \bibnamefont {Cremon}},\ and\ \bibinfo {author} {\bibfnamefont {S.~M.}\ \bibnamefont {Reimann}},\ }\bibfield  {title} {\bibinfo {title} {Erratum: Ground-state properties of few dipolar bosons in a quasi-one-dimensional harmonic trap [phys. rev. a 81, 063616 (2010)]},\ }\href {https://doi.org/10.1103/PhysRevA.87.039903} {\bibfield  {journal} {\bibinfo  {journal} {Phys. Rev. A}\ }\textbf {\bibinfo {volume} {87}},\ \bibinfo {pages} {039903} (\bibinfo {year} {2013})}\BibitemShut {NoStop}%
\bibitem [{Note1()}]{Note1}%
  \BibitemOpen
  \bibinfo {note} {We define the term condensate as referring to the $k=0$ mode.}\BibitemShut {Stop}%
\bibitem [{\citenamefont {Leggett}(2001)}]{Leggett_2001}%
  \BibitemOpen
  \bibfield  {author} {\bibinfo {author} {\bibfnamefont {A.~J.}\ \bibnamefont {Leggett}},\ }\bibfield  {title} {\bibinfo {title} {Bose-einstein condensation in the alkali gases: Some fundamental concepts},\ }\href {https://doi.org/10.1103/RevModPhys.73.307} {\bibfield  {journal} {\bibinfo  {journal} {Rev. Mod. Phys.}\ }\textbf {\bibinfo {volume} {73}},\ \bibinfo {pages} {307} (\bibinfo {year} {2001})}\BibitemShut {NoStop}%
\bibitem [{Note2()}]{Note2}%
  \BibitemOpen
  \bibinfo {note} {By {\protect \it ab initio} calculation, we mean to find a direct solution of the model, which is many-body but already one dimensional and with effective potentials for dipolar and short-range interactions.}\BibitemShut {Stop}%
\bibitem [{\citenamefont {Dutta}\ \emph {et~al.}(2022)\citenamefont {Dutta}, \citenamefont {Buyskikh}, \citenamefont {Daley},\ and\ \citenamefont {Mueller}}]{Dutta2022Jun}%
  \BibitemOpen
  \bibfield  {author} {\bibinfo {author} {\bibfnamefont {S.}~\bibnamefont {Dutta}}, \bibinfo {author} {\bibfnamefont {A.}~\bibnamefont {Buyskikh}}, \bibinfo {author} {\bibfnamefont {A.~J.}\ \bibnamefont {Daley}},\ and\ \bibinfo {author} {\bibfnamefont {E.~J.}\ \bibnamefont {Mueller}},\ }\bibfield  {title} {\bibinfo {title} {{Density Matrix Renormalization Group for Continuous Quantum Systems}},\ }\href {https://doi.org/10.1103/PhysRevLett.128.230401} {\bibfield  {journal} {\bibinfo  {journal} {Phys. Rev. Lett.}\ }\textbf {\bibinfo {volume} {128}},\ \bibinfo {pages} {230401} (\bibinfo {year} {2022})}\BibitemShut {NoStop}%
\bibitem [{Note3()}]{Note3}%
  \BibitemOpen
  \bibinfo {note} {See Supplemental Material for details.}\BibitemShut {Stop}%
\bibitem [{\citenamefont {Li}\ \emph {et~al.}(2023{\natexlab{b}})\citenamefont {Li}, \citenamefont {Zhang}, \citenamefont {Yang}, \citenamefont {Lin}, \citenamefont {Gopalakrishnan}, \citenamefont {Rigol},\ and\ \citenamefont {Lev}}]{Li2023Jun}%
  \BibitemOpen
  \bibfield  {author} {\bibinfo {author} {\bibfnamefont {K.-Y.}\ \bibnamefont {Li}}, \bibinfo {author} {\bibfnamefont {Y.}~\bibnamefont {Zhang}}, \bibinfo {author} {\bibfnamefont {K.}~\bibnamefont {Yang}}, \bibinfo {author} {\bibfnamefont {K.-Y.}\ \bibnamefont {Lin}}, \bibinfo {author} {\bibfnamefont {S.}~\bibnamefont {Gopalakrishnan}}, \bibinfo {author} {\bibfnamefont {M.}~\bibnamefont {Rigol}},\ and\ \bibinfo {author} {\bibfnamefont {B.~L.}\ \bibnamefont {Lev}},\ }\bibfield  {title} {\bibinfo {title} {{Rapidity and momentum distributions of one-dimensional dipolar quantum gases}},\ }\href {https://doi.org/10.1103/PhysRevA.107.L061302} {\bibfield  {journal} {\bibinfo  {journal} {Phys. Rev. A}\ }\textbf {\bibinfo {volume} {107}},\ \bibinfo {pages} {L061302} (\bibinfo {year} {2023}{\natexlab{b}})}\BibitemShut {NoStop}%
\bibitem [{\citenamefont {Yang}\ \emph {et~al.}(2023)\citenamefont {Yang}, \citenamefont {Zhang}, \citenamefont {Li}, \citenamefont {Lin}, \citenamefont {Gopalakrishnan}, \citenamefont {Rigol},\ and\ \citenamefont {Lev}}]{Yang2023Aug}%
  \BibitemOpen
  \bibfield  {author} {\bibinfo {author} {\bibfnamefont {K.}~\bibnamefont {Yang}}, \bibinfo {author} {\bibfnamefont {Y.}~\bibnamefont {Zhang}}, \bibinfo {author} {\bibfnamefont {K.-Y.}\ \bibnamefont {Li}}, \bibinfo {author} {\bibfnamefont {K.-Y.}\ \bibnamefont {Lin}}, \bibinfo {author} {\bibfnamefont {S.}~\bibnamefont {Gopalakrishnan}}, \bibinfo {author} {\bibfnamefont {M.}~\bibnamefont {Rigol}},\ and\ \bibinfo {author} {\bibfnamefont {B.~L.}\ \bibnamefont {Lev}},\ }\bibfield  {title} {\bibinfo {title} {{Phantom energy in the nonlinear response of a quantum many-body scar state}},\ }\bibfield  {journal} {\bibinfo  {journal} {arXiv}\ }\href {https://doi.org/10.48550/arXiv.2308.11615} {10.48550/arXiv.2308.11615} (\bibinfo {year} {2023}),\ \Eprint {https://arxiv.org/abs/2308.11615} {2308.11615} \BibitemShut {NoStop}%
\bibitem [{\citenamefont {Haller}\ \emph {et~al.}(2009)\citenamefont {Haller}, \citenamefont {Gustavsson}, \citenamefont {Mark}, \citenamefont {Danzl}, \citenamefont {Hart}, \citenamefont {Pupillo},\ and\ \citenamefont {N{\ifmmode\ddot{a}\else\"{a}\fi}gerl}}]{Haller2009Sep}%
  \BibitemOpen
  \bibfield  {author} {\bibinfo {author} {\bibfnamefont {E.}~\bibnamefont {Haller}}, \bibinfo {author} {\bibfnamefont {M.}~\bibnamefont {Gustavsson}}, \bibinfo {author} {\bibfnamefont {M.~J.}\ \bibnamefont {Mark}}, \bibinfo {author} {\bibfnamefont {J.~G.}\ \bibnamefont {Danzl}}, \bibinfo {author} {\bibfnamefont {R.}~\bibnamefont {Hart}}, \bibinfo {author} {\bibfnamefont {G.}~\bibnamefont {Pupillo}},\ and\ \bibinfo {author} {\bibfnamefont {H.-C.}\ \bibnamefont {N{\ifmmode\ddot{a}\else\"{a}\fi}gerl}},\ }\bibfield  {title} {\bibinfo {title} {{Realization of an Excited, Strongly Correlated Quantum Gas Phase}},\ }\href {https://doi.org/10.1126/science.1175850} {\bibfield  {journal} {\bibinfo  {journal} {Science}\ }\textbf {\bibinfo {volume} {325}},\ \bibinfo {pages} {1224} (\bibinfo {year} {2009})}\BibitemShut {NoStop}%
\bibitem [{\citenamefont {Zin}\ \emph {et~al.}(2021)\citenamefont {Zin}, \citenamefont {Pylak}, \citenamefont {Wasak}, \citenamefont {Jachymski},\ and\ \citenamefont {Idziaszek}}]{Zin2021Sep}%
  \BibitemOpen
  \bibfield  {author} {\bibinfo {author} {\bibfnamefont {P.}~\bibnamefont {Zin}}, \bibinfo {author} {\bibfnamefont {M.}~\bibnamefont {Pylak}}, \bibinfo {author} {\bibfnamefont {T.}~\bibnamefont {Wasak}}, \bibinfo {author} {\bibfnamefont {K.}~\bibnamefont {Jachymski}},\ and\ \bibinfo {author} {\bibfnamefont {Z.}~\bibnamefont {Idziaszek}},\ }\bibfield  {title} {\bibinfo {title} {{Quantum droplets in a dipolar Bose gas at a dimensional crossover}},\ }\href {https://doi.org/10.1088/1361-6455/ac2244} {\bibfield  {journal} {\bibinfo  {journal} {J. Phys. B: At. Mol. Opt. Phys.}\ }\textbf {\bibinfo {volume} {54}},\ \bibinfo {pages} {165302} (\bibinfo {year} {2021})}\BibitemShut {NoStop}%
\bibitem [{\citenamefont {Pylak}\ \emph {et~al.}(2024)\citenamefont {Pylak}, \citenamefont {Gajda},\ and\ \citenamefont {Zin}}]{Pylak2024}%
  \BibitemOpen
  \bibfield  {author} {\bibinfo {author} {\bibfnamefont {M.}~\bibnamefont {Pylak}}, \bibinfo {author} {\bibfnamefont {M.}~\bibnamefont {Gajda}},\ and\ \bibinfo {author} {\bibfnamefont {P.}~\bibnamefont {Zin}},\ }\bibfield  {title} {\bibinfo {title} {{Dipolar Droplets at 3D-1D Crossover}},\ }\bibfield  {journal} {\bibinfo  {journal} {arXiv}\ }\href {https://doi.org/10.48550/arXiv.2405.15433} {10.48550/arXiv.2405.15433} (\bibinfo {year} {2024}),\ \Eprint {https://arxiv.org/abs/2405.15433} {2405.15433} \BibitemShut {NoStop}%
\end{thebibliography}%

\end{document}

\preprint{APS/123-QED}

\title{Supplemental Material for:\\ Restored condensate fraction and correlated droplets in one-dimensional dipolar Bose gas}

\author{Bu\u{g}ra T\"uzemen}
\email{btuzemen@ifpan.edu.pl}
\affiliation{
Institute of Physics, Polish Academy of Sciences, Al. Lotnik\'{o}w 32/46, 02-668 Warsaw, Poland
}
\affiliation{
Center for Theoretical Physics, Polish Academy of Sciences, Al. Lotnik\'{o}w 32/46, 02-668 Warsaw, Poland
}

\author{Maciej Marciniak}
\thanks{Bu\u{g}ra T\"uzemen and Maciej Marciniak contributed equally to this work.}
\affiliation{
Center for Theoretical Physics, Polish Academy of Sciences, Al. Lotnik\'{o}w 32/46, 02-668 Warsaw, Poland
}

\author{Krzysztof Paw\l owski}
\affiliation{
Center for Theoretical Physics, Polish Academy of Sciences, Al. Lotnik\'{o}w 32/46, 02-668 Warsaw, Poland
}

\date{\today}

\begin{abstract}
The technical aspects of ab initio many-body methods are discussed, with detailed explanations of the calculations for the first and second-order correlation functions, as well as the number-conserving Bogoliubov approach.
\end{abstract}

\maketitle

\setcounter{figure}{5}   
\setcounter{equation}{3}

\subsection{Ab initio many-body methods} \label{sec:Appendix_cDMRG}

In this study, a significant portion of the data comes from ab initio many-body calculations. These calculations involve determining the ground state of the Hamiltonian, which governs a system of interacting bosons confined in a box of length $L$ with periodic boundary conditions. In the framework of the second quantization, the Hamiltonian operator is represented as:
\begin{align} \label{eq:AppDMRG_H}
    \hat{H} = \frac{\hbar^2}{2m} \int_0^L dx \frac{\partial \hat{\Psi}^{\dagger}(x)}{\partial x} \frac{\partial \hat{\Psi}(x)}{\partial x}
    + \iint_0^L dx dx' V_{ai}(x-x') \hat{\Psi}^{\dagger}(x)\hat{\Psi}^{\dagger}(x')\hat{\Psi}(x')\hat{\Psi}(x),
\end{align}

where interparticle potential $V_{ai}(x-x') = g\delta(x-x')+g_{dd}v^{\sigma}_{dd}(x-x')$ and $\hat{\Psi}(x)= \sum_{n} f_{n}(x) \hat{b}_n$ is the bosonic operator annihilating the particle in position $x$. It is important to note that the structure of the Hamiltonian matrix varies based on the chosen basis of single-body states used to expand the field operators. The selection of an appropriate basis depends heavily on the system. For our investigation, particularly for small systems with $N \leq 4$, we employed the plane wave basis, where the single-particle states are given by $f_n(x) = L^{-1/2}e^{i2\pi n x/L}$. This choice results in the Hamiltonian matrix taking the form:

\begin{align} \label{eq:PlaneWavesHamiltonian}
    \hat{H}_{pw} = \frac{2\pi\hbar^2}{m L^2}\sum_{j=-\infty}^{\infty} j^2 \hat{n}_j + \frac{1}{2L}\sum_{j,j',k}  V_{k}\hat{b}^{\dagger}_{j-k} \hat{b}^{\dagger}_{j'+k} \hat{b}_{j'} \hat{b}_{j},
\end{align}
where {\color{red}$V_{k}$} is a Fourier transform of the interaction potential.

The computational complexity associated with finding the ground state of such Hamiltonian increases rapidly as the number of atoms grows. Therefore, for systems with $N > 4$, it is advantageous to use an alternative technique: the Density Matrix Renormalization Group for Continuous Quantum Systems (cDMRG) \cite{Dutta2022Jun}, adapted for systems with nonlocal interactions. This method involves dividing the box containing the atoms into smaller subspaces and expanding the field operators into a Taylor series. In practice, this entails introducing an orthonormal basis of single-particle states in each subspace, represented by Laguerre polynomials. Since each single-particle state is defined by two quantum numbers—the subspace where the orbital is located and the order of the Laguerre polynomial—it is convenient to rewrite the annihilation field operator as $\hat{\Psi}(x) = \sum_{j,n} f_{j,n}(x) \hat{b}_{j,n}$, with the explicit form of the single-particle states given by:

\begin{align} \label{eq:LaguerrePolynomials}
     \begin{cases}
        f_{j,0}(x) = \frac{1}{\sqrt{\Delta}}\square_j(x) = g_{j,0}(x)\square_j(x),
        \\
        f_{j,1}(x) =\frac{ \sqrt{3}}{\sqrt{\Delta}}(2\frac{x}{\Delta}-(2j+1)) \square_j(x) = g_{j,1}(x)\square_j(x),
        \\
        f_{j,2}(x) = \frac{\sqrt{5}}{2\sqrt{\Delta}}(3(2\frac{x}{\Delta}-(2j+1))^2-1) \square_j(x) = g_{j,2}(x)\square_j(x)\\
    \end{cases}
\end{align}
where the symbol $\square$ is defined as
\begin{equation}
    \square_j(x) = 
\begin{cases} 
1, & \quad\text{if } j\Delta \leq x < (j+1)\Delta \\
0, & \text{elsewhere},
\end{cases}
\end{equation}
with $\Delta$ being the length of the subspace interval.
Adopting such a basis leads to the Hamiltonian Eq.~\ref{eq:Hsc}
\begin{align} \begin{split}\label{eq:Hsc}
    \hat{H}_{lp}  &=  \frah \sum_{j,\bm{n_2}} \bigg[K_{\bm{n_2}}
    +
    \frac{g}{2}\sum_{j,\bm{n_4}} U_{\bm{n_4}}\hbd{j,n}\hbd{j,n'}\hb{j,n''}\hb{j,n'''}
    +
    \frac{g_{dd}}{2}\sum_{j\leq j',\bm{n_4}} U^{dd}_{d,\bm{n_4}}\hbd{j,n}\hbd{j',n'}\hb{j',n''}\hb{j,n'''}
    \\&+
    \sum_{j,\bm{n_2}}\bigg[
    \big( \Lambda \Upsilon^{L}_{0,\bm{n_2}} +\Upsilon^{c}_{0,\bm{n_2}}\big)\hbd{j,n}\hb{j,n'} 
    +
    \big( \Lambda \Upsilon^{L}_{1,\bm{n_2}} +\Upsilon^{c}_{1,\bm{n_2}}\big)\hbd{j+1,n}\hb{j,n'} 
    +
    \big( \Lambda \Upsilon^{L}_{-1,\bm{n_2}} +\Upsilon^{c}_{-1,\bm{n_2}}\big)\hbd{j,n}\hb{j+1,n'} 
    \bigg]
\end{split}\end{align}
where $\bm{n_2}=n,n'$, $\bm{n_4}=n,n',n'',n'''$ and the values of a matrix elements $K$,$U$, $U_{dd}$ and $\Upsilon$ are given by the equations below:
\begin{align} 
    K_{\bm{n_2}} = \int_{0}^{\Delta} h^*_{0,n}(x) h_{0,n'}(x)  \dx,
\end{align}
\begin{align}
    U_{\bm{n_4}} = \int_{0}^{\Delta} g_{0,n}(x) g_{0,n'}(x)g_{0,n''}(x) g_{0,n'''}(x)dx,
\end{align}
\begin{align}
\begin{aligned}
    \Upsilon_{0,\bm{n_2}}^L &= g_{0,n}(0)g_{0,n'}(0) 
    + g_{0,n}(\Delta)g_{0,n'}(\Delta) \\
    \Upsilon_{1,\bm{n_2}}^L &= g_{0,n}(0)g_{0,n'}(\Delta) \\
    \Upsilon_{-1,\bm{n_2}}^L &= -g_{0,n}(\Delta)g_{0,n'}(0) \\
    \Upsilon_{0,\bm{n_2}}^c &= h_{0,n}(0)g_{0,n'}(0) + g_{0,n}(0)h_{0,n'}(0) - h_{0,n}(\Delta)g_{0,n'}(\Delta) - g_{0,n}(\Delta)h_{0,n'}(\Delta) \\
    \Upsilon_{1,\bm{n_2}}^c &= -h_{0,n}(0)g_{0,n'}(\Delta) + g_{0,n}(0)h_{0,n'}(\Delta) \\
    \Upsilon_{-1,\bm{n_2}}^c &= h_{0,n}(\Delta)g_{0,n'}(0) - g_{0,n}(\Delta)h_{0,n'}(0)
\end{aligned}
\end{align}
and 
\begin{align}
\tilde{U}^{dd}_{d,\bm{n}} = \int_{S_1(r)} v_{dd}^{\sigma}(r) \bigg[\int_{S_1(R)} \gamma_{d,\bm{n}}(r,R)  dR \bigg] dr 
+ \int_{S_2(r)} v_{dd}^{\sigma}(r) \bigg[\int_{S_2(R)} \gamma_{d,\bm{n}}(r,R)  dR \bigg] dr,
\end{align}

where $\gamma_{d,\bm{n}}(r,R) = g_{0,n}(R-r/2) g_{d,n'}(R+r/2) g_{d,n''}(R+r/2) g_{0,n'''}(R-r/2)$ and integration areas are given as
$S_1(r,R) = \big\{ (r,R): (d-1)\Delta \leq r \leq d\Delta \;\;\land\;\; d\Delta-r/2 \leq R\leq \Delta+r/2\big\}$ and $S_2(r,R) = \big\{ (r,R): d\Delta \leq r \leq (d+1)\Delta \;\;\land\;\; r/2\leq R\leq (d+1)\Delta-r/2\big\}$.
The definition of the virtual parameter $\Lambda$ is well explained in \cite{Dutta2022Jun}.
A significant advantage of this Hamiltonian formulation is that for interatomic interactions that decay rapidly with distance—such as dipole interactions between atoms in a tightly confined trap—the matrix elements of the Hamiltonian for distant nodes become negligible. This enables the effective application of methods like DMRG to study the system’s ground state. In our work, all data for systems with $N > 4$ 
  obtained using ab initio methods were computed using the iTensor library \cite{itensor,itensor-r0.3}.

\begin{figure}[t]
\includegraphics[width=8cm]{cDMRG_1.drawio.png}
\centering
\caption{A demonstration of the first three single-particle states. It can be seen that all of them are localized in the same spacial subspace (indexed by $j=1$), but they are characterized by different "degree" quantum numbers $n$. \label{fig:cDMRG_basis}}
\end{figure}

\subsection{Convergence analysis}
For the selected physical parameters of the system—such as the number of atoms, spatial dimensions of the box, and interactions—the following numerical parameters can be adjusted: (1) local cutoff $n_c$, (2) number of subspaces $N_s$, (3) cutoff for the range of nonlocal interactions $r$, and (4) the parameter $\Lambda$ in the final cycle, which determines the continuity of the wavefunction. In the calculations presented in this manuscript, we consistently assumed the local cutoff to be equal to the number of atoms, $n_c=N$. In Fig.~\ref{fig:Convergence_cDMRG}, we provide a plot showing the convergence analysis for the system discussed in the main body of the text.

\begin{figure}[t]
\includegraphics[width=\columnwidth]{Convergence_cDMRG_momentumDistribution_v4.png}
\centering
\caption{We present the correlation functions (top row) and the corresponding momentum distributions (bottom row) obtained numerically using cDMRG. The results are shown for various numerical parameters, including the number of subspaces used $N_s$ and, when applicable, different cutoffs for dipolar interactions $r$. For $\theta=\pi/2$ and $\theta = 55^{\circ}$ (where dipolar interactions can be neglected), the correlation function plots are supplemented with insets that provide a more detailed view of the correlation functions near the point where $n x \to 3$.\label{fig:Convergence_cDMRG}}
\end{figure}
%%%%%%%%%%%%%%%%%%%%%%%%%%%%%%%%%%
\subsection{Correlation functions}
%%%%%%%%%%%%%%%%%%%%%%%%%%%%%%%%%%
We are particularly interested in the first and second-order correlation functions, defined as 
\begin{align}
G_1(x,x') &= \int_0^L \Psi^*(x,x_2,\dots, x_N) \Psi(x',x_2,\dots,x_N) \, \mathrm{d}x_2 \dots \mathrm{d}x_N \notag \\
&= \langle \hat{\Psi}^\dagger(x)\hat{\Psi}(x')\rangle,
\end{align}
and 
\begin{align}
G_2(x,x') &= \int_0^L |\Psi(x,x',\dots,x_N)|^2 \, \mathrm{d}x_3 \dots \mathrm{d}x_N \notag \\
&= \langle \hat{\Psi}^\dagger(x)\hat{\Psi}^\dagger(x')\hat{\Psi}(x')\hat{\Psi}(x)\rangle.
\end{align}
These formulas, in the basis used in the cDMRG approach, can be explicitly expanded as:
\begin{align} \label{eq:cDMRG_correlationFunctions}
    G_1(x,x') &= \sum_{n,n'}g_{j_{(x)},n}(x)g_{j_{(x')},n'}(x')\langle \hbd{j(x),n}\hb{j(x'),n'}\rangle,
    \\G_2(x,x') &= \sum_{\bm{n_4}}g_{j_{(x)},n}(x) g_{j_{(x')},n'}(x') g_{j_{(x')},n''}(x') g_{j_{(x)},n'''}(x)\langle\hbd{j(x),n}\hbd{j(x'),n'}\hb{j(x'),n''}\hb{j(x),n'''}\rangle.
\end{align}
where function $j(x)$ maps the position to the index of a corresponding subspace, i.e. $j(x) = \rm{floor}(x/\Delta)$. Then, as the second-order correlations are interested in themselves to study the bunching and antibunching properties of a system, the first-order correlation function can be used to investigate the momentum distribution:
\begin{align} \label{eq:MomentumDistribution}
    n(p) = \iint_0^L  e^{ip(x-x')/\hbar} G_1(x,x')\rm{dxdx'}.
\end{align}
and the quantum depletion  $\delta N$ which for a system with periodic boundary conditions, in the cases without self-bound states, is:
\begin{align} \label{eq:CondensateFraction}
    \delta N/N = 1-\frac{1}{N}\iint_0^L G_1(x,x') \rm{dxdx'}.
\end{align}

\subsection{Number-conserving Bogoliubov method}

The many-body Hamiltonian that describes the interacting ultracold bosons is:
\begin{equation}
\mathcal{H} = \sum_{k} h_{k} \adk{k} \ak{k} + \frac{1}{2L}\sum_{k,k^\prime,q} V_{q} \adk{k+q}\adkp{k-q}\akp{k}\ak{k}. 
\end{equation}

The Bogoliubov method assumes the number of depleted particles is negligible compared to the number of particles in the $k=0$ mode. Therefore, it disregards the interaction between the depleted particles. As a result, the effective Hamiltonian is:

\begin{align}
    \mathcal{H}_0 = \frac{1}{2L} V_0 \hat{n}_0(\hat{n}_0-1) + \sum_{k} h_{k} \hat{n}_k 
    + \frac{1}{L} \sumk (V_0+V_k) \hat{n}_k \hat{n}_{0} + \frac{1}{2L}\sumk V_k \bigg[\adk{k}\adk{-k}\ak{0}\ak{0} + h.c.\bigg],
\end{align}
where $\hat{n}_k = \adk{k}\ak{k}$.
To compute the expectation value of the above Hamiltonian, We use the following ground-state ansatz~\cite{Leggett_2001}:

\begin{equation}
\ket{\Psi_{\mathrm{BdG}}} = C \left( \hat{a}_0^{\dagger}\hat{a}_0^{\dagger} - \sum_{k \neq 0}^{\infty} \frac{v_k}{u_k} \hat{a}_k^\dagger \hat{a}_{-k}^\dagger \right) ^{N/2} \ket{{\rm vacuum}}.  
\label{eq:bdg_ansatz}
\end{equation}

Here, $|u_{k}|^2 = 1 + |v_{k}|^2 = (1/2)\left[ \left(h_{k} + n V_{k}\right)/\varepsilon_{k} + 1 \right]$, are the Bogoliubov amplitudes, $\varepsilon_{k} = \sqrt{ h_{k}\left(h_{k}+2nV_{k} \right)}$ is the Bogoliubov eigenvalue,
and $C$ is a normalization constant. Since we do not treat the depleted states exactly, we employ the highest-order Taylor series expansion of the wave function:

\begin{equation}
    \ket{\Psi^{(0)}_{\mathrm{BdG}}} = C \bigg(\sqrt{N!}\ket{0:N}-\frac{1}{2}N \sqrt{(N-2)!} \sum_{k\neq 0}^{\infty} \frac{v_{k}}{u_{k}}\ket{0:N-2,-k:1,k:1}\bigg),
\end{equation}
where $\ket{0:N}$ and $\ket{0:N-2,-k:1,k:1}$ correspond to Fock states where all atoms occupy the single particle state with momentum equals zero and to Fock state where one atom occupy mode with momentum $-k$, one with momentum $k$ and $N-2$ atoms occupy the mode with momentum equals zero. As a result, the ground-state energy is:

\begin{eqnarray}
E_{\mathrm{GS}} &=&  \bra{\Psi^{(0)}_{\mathrm{BdG}}}\mathcal{H}_{0}\ket{\Psi^{(0)}_{\mathrm{BdG}}} \nonumber \\
&=& \frac{1}{2}V_0 \left[ N(N-1) |C_0|^2 + (N-2)(N-3)(1 - |C_0|^2) \right] \nonumber \\
&+& 2 \sum_{k>0} \left( \left(h_k + (N-2)(V_{k} + V_0) \right) |C_k|^2 \right) \nonumber \\
&+& 2 \sum_{k>0} \sqrt{N(N-1)} C_k^* C_0 V_k,
\end{eqnarray}
where $C_{0} = C\sqrt{N!}/L$, and $C_{k} = -\frac{v_{k}}{2u_{k}}CN\sqrt{\left(N-2\right)!}/L$.

\bibliography{biblio}